\def\cross{\bds\times}

\def\nddLII{\overset{\cdot}{\Omega}_{.2}^\mathrm{L\,II}\,}
\def\nddL{\overset{\cdot}{\Omega}_{.2}^\mathrm{L}\,}
\def\ndqLII{\overset{\cdot}{\Omega}_{.4}^\mathrm{L\,II}\,}
\def\ndqL{\overset{\cdot}{\Omega}_{.4}^\mathrm{L}\,}
\def\ndsLII{\overset{\cdot}{\Omega}_{.6}^\mathrm{L\,II}\,}
\def\ndsL{\overset{\cdot}{\Omega}_{.6}^\mathrm{L}\,}
\def\etdLII{\overset{\cdot}{\eta}_{.2}^\mathrm{L\,II}\,}
\def\etdL{\overset{\cdot}{\eta}_{.2}^\mathrm{L}\,}
\def\etqLII{\overset{\cdot}{\eta}_{.4}^\mathrm{L\,II}\,}
\def\etqL{\overset{\cdot}{\eta}_{.4}^\mathrm{L}\,}
\def\etsLII{\overset{\cdot}{\eta}_{.6}^\mathrm{L\,II}\,}
\def\etsL{\overset{\cdot}{\eta}_{.6}^\mathrm{L}\,}
\def\cross{\bds\times}

\def\nk{n_{\rm b}}

\def\Pb{P_{\rm b}}

\def\rfr#1{Equation~(\ref{#1})}
\def\rfrs#1#2{Equations~(\ref{#1})~to~(\ref{#2})}

\def\derp#1#2{\rp{\partial{#1}}{\partial{#2}}}
\def\dert#1#2{\frac{{{\textrm{d}}}{#1}}{{{\textrm{d}}}{#2}}}
\def\virg#1{``#1"}

\def\eqi{\begin{equation}}
\def\eqf{\end{equation}}
\def\eqia{\begin{eqnarray}}
\def\eqfa{\end{eqnarray}}

\def\rp#1#2{{#1\over#2}}
\def\lb#1{\label{#1}}
\def\kap{\bds{\hat{S}}}

\def\bds#1{\boldsymbol{#1}}
\def\ton#1{\left(#1\right)}
\def\qua#1{\left[#1\right]}
\def\grf#1{\left\{#1\right\}}
\def\ang#1{\left\langle #1\right\rangle}
\documentclass[onecolumn]{aastex}
\usepackage{morefloats}
\usepackage[title]{appendix}
\usepackage{hyperref}
\usepackage{booktabs}
\usepackage[table,xcdraw]{xcolor}
\usepackage{multirow}
\usepackage{rotating,tabularx}
\usepackage{float}
\usepackage{enumerate}
\usepackage{rotating}
\usepackage[polutonikogreek,english]{babel}
\usepackage{amsmath,starfont,textgreek,w-greek,wasysym}
\usepackage{amsthm}
\usepackage{amscd,lineno}
\usepackage{amssymb,dsfont}
\usepackage{graphicx,epsfig}
\usepackage{txfonts}
\usepackage{xr-hyper}
\RequirePackage{color}

\newcommand{\emaila}{lorenzo.iorio@libero.it}

\allowdisplaybreaks[1]

\begin{document}

\title{On the mean anomaly and the mean longitude in tests of post-Newtonian gravity}

\shortauthors{L. Iorio}

\author{Lorenzo Iorio\altaffilmark{1} }
\affil{Ministero dell'Istruzione, dell'Universit\`{a} e della Ricerca
(M.I.U.R.)-Istruzione
\\ Permanent address for correspondence: Viale Unit\`{a} di Italia 68, 70125, Bari (BA),
Italy}

\email{\emaila}

\begin{abstract}
The distinction between the mean anomaly $\mathcal{M}(t)$ and the mean anomaly at epoch $\eta$, and the mean longitude $l(t)$ and the mean longitude at epoch $\epsilon$ is clarified in the context of a their possible use in post-Keplerian tests of gravity, both Newtonian and post-Newtonian. In particular, the perturbations induced on $\mathcal{M}(t),\,\eta,\,l(t),\,\epsilon$ by the post-Newtonian Schwarzschild and Lense-Thirring fields, and the classical accelerations due to the atmospheric drag and the oblateness $J_2$ of the central body are calculated for an arbitrary orbital configuration of the test particle and a generic orientation of the primary's spin axis $\bds{\hat{S}}$. They provide us with further observables which could be fruitfully used, e.g., in  better characterizing astrophysical binary systems and in more accurate satellite-based tests around major bodies of the Solar System. Some erroneous claims by Ciufolini and Pavlis appeared in the literature are confuted. In particular, it is shown that there are no net perturbations of the Lense-Thirring acceleration on either the semimajor axis $a$ and the mean motion $\nk$. Furthermore, the quadratic signatures on $\mathcal{M}(t)$ and $l(t)$ due to certain disturbing non-gravitational accelerations like the atmospheric drag can be effectively disentangled from the post-Newtonian linear trends of interest provided that a sufficiently long temporal interval for the data analysis is assumed. A possible use of $\eta$ along with the longitudes of the ascending node $\Omega$ in tests of general relativity with the existing LAGEOS and LAGEOS II satellites is suggested.
\end{abstract}

keywords{
General relativity and gravitation; Experimental studies of gravity; Experimental
tests of gravitational theories; Satellite orbits
}
\section{Introduction}
In regard to possible tests of post-Newtonian (pN) features of general relativity\footnote{For a recent overview of the current status and challenges of the Einsteinian theory of gravitation, see, e.g., Debono and Smoot \cite{2016Univ....2...23D}, and references therein.} and of alternative models of gravity with, e.g., Earth's artificial satellites, Solar system's planets and other astrophysical binaries, there  is a considerable confusion in the literature about the possible use of the mean anomaly $\mathcal{M}\ton{t}$ as potential observable in addition to the widely inspected argument of pericentre $\omega$ and, to a lesser extent, longitude of the ascending node $\Omega$. Indeed, it is as if some researchers, including the present author,  who have tried to compute perturbatively the mean rate of change of the mean anomaly in excess with respect to the Keplerian case due to some pN accelerations were either unaware of the fact that what they, actually, calculated was the secular precession of the mean anomaly at the epoch $\eta$, or they systematically neglected a potentially non-negligible contribution to the overall change of the mean anomaly induced indirectly by the semimajor axis $a$ through the mean motion $\nk$. Such a confusion has produced so far some misunderstanding which led, e.g., to unfounded criticisms about alleged proposals of using the mean anomaly, especially in the case of man-made spacecraft orbiting the Earth, or even uncorrect evaluations of the total pN effects sought. An example that sums up well the aforementioned confusion and misunderstanding, even in the peer-reviewed literature, is the following one. Ciufolini and Pavlis \cite{2005NewA...10..636C} wrote \virg{[\ldots] one of the most profound mistakes and misunderstandings of Iorio (2005) is the proposed use of the mean anomaly of a satellite to measure the Lense-Thirring effect [\ldots] This is simply a nonsense statement: let us, for example, consider a satellite at the LAGEOS altitude, the Lense-Thirring effect on its mean longitude is of the order of 2 m/y, however, the mean longitude change is about $1.8\times 10^{11}$ m/y. Thus, from Kepler's law, the Lense-Thirring effect corresponds to a change of the LAGEOS semi-major of less than $0.009$ cm! Since, even a high altitude satellite such as LAGEOS showed a semimajor axis change of the order of 1 mm/day, due to atmospheric drag and to the Yarkoski-Rubincam effect (because of atmospheric drag, the change of semimajor axis and mean motion is obviously
much larger for lower altitude satellites), and since the present day precision of satellite laser ranging is, even in the case of the best SLR stations, of several millimeters, it is a clear nonsense to propose a test of the Lense-Thirring effect based on using the mean anomaly of \textit{any} satellite, mean anomaly largely affected by non-conservative forces.} It is difficult to understand what is the target of the arrows by Ciufolini and Pavlis \cite{2005NewA...10..636C} since the mean anomaly is not even mentioned in the published version of the criticized paper by the present author, not to mention any explicitly detailed proposal to use it.  Be that as it may, in the following, we will show that, actually, using the mean anomaly, or the mean longitude $l\ton{t}$, in pN tests with artificial Earth's satellites may be feasible,  provided that certain non-gravitational perturbations are compensated by some active drag-free mechanism. However, even in case of passive, geodetic satellites, we will show that, under certain conditions, it is possible to  separate the relativistic linear trends of interest from the unwanted parabolic signatures of non-conservative origin. Furthermore, the arguments provided by Ciufolini and Pavlis \cite{2005NewA...10..636C} about the Lense-Thirring effect and the mean longitude are erroneous. Finally, the use of the mean anomaly at epoch  or of the mean longitude at epoch $\epsilon$ is, in principle, possible even with passive, geodetic spacecraft like those of the LAGEOS family  because they are, by construction, free from the aforementioned potential drawbacks exhibited by the mean anomaly and the mean longitude themselves, which was completely ignored or unrecognized by Ciufolini and Pavlis \cite{2005NewA...10..636C}.

The paper is organized as follows. In Section\,\ref{theo}, we will review the basics of the mean anomaly, the mean anomaly at epoch, the mean longitude, and the mean longitude at epoch along with the calculation of their perturbations with respect to the purely Keplerian case in presence of a generic disturbing post-Keplerian (pK) acceleration. Section\,\ref{Greffects} is devoted to the calculation of the effects of some well-known pN accelerations (Schwarzschild and Lense-Thirring), while the impact of the atmospheric drag and the oblateness of the primary are treated in Section\,\ref{Oblo}. The potential of a possible use of the mean anomaly at epoch in the ongoing tests with the satellites LAGEOS and LAGEOS II is discussed in Section\,\ref{combos}. Section\,\ref{fine} summarizes our findings, and offers our conclusions.
\section{The mean anomaly and the mean longitude}\lb{theo}
\subsection{The mean anomaly}
In the restricted two-body problem, the mean anomaly $\mathcal{M}(t)$ is one of the three time-dependent fast angular variables which, in celestial mechanics, can be used to characterize the instantaneous position of a test particle along its Keplerian ellipse, being the eccentric anomaly $E$ and the true anomaly $f$ the other  two anomalies. In the unperturbed Keplerian case, the mean anomaly is defined as
\eqi
\mathcal{M}(t) \doteq \eta + \nk\,\ton{t-t_0},
\eqf
where\footnote{The symbol $\eta$ is used for the mean anomaly at epoch by Milani et al. \cite{Nobilibook87}. In the notation by Brumberg \cite{1991ercm.book.....B}, the mean anomaly is $l$, while the mean anomaly at epoch is $l_0$. Kopeikin et al. \cite{2011rcms.book.....K} denote $\eta$ as $\mathcal{M}_0$, while Bertotti et al. \cite{Bertotti03} adopt $\epsilon^{'}$.} $\eta$ is the mean anomaly at the reference epoch $t_0$, and \eqi
\nk=\sqrt{\rp{\mu}{a^3}}\lb{Kep}
\eqf is the Keplerian mean motion. In \rfr{Kep}, $\mu\doteq GM$ is the gravitational parameter of the primary having mass $M$, while $G$ is the Newtonian constant of gravitation; in the following, we will assume $\mu=\textrm{const}$. The mean anomaly at epoch $\eta$ is  one of the six Keplerian orbital elements parameterizing the orbit of a test particle in space. In the unperturbed case, $\mathcal{M}(t)$ is a linear function of time $t$ because both $a$ and $\eta$ are constants of motion. If a relatively small perturbing pK acceleration $\bds A$ is present, both $a$ and $\eta$ are, in general, affected by it, becoming time-dependent. As a result, also the mean motion is, in general, modified so that
\eqi
\nk\rightarrow\nk^\textrm{pert} = \nk + \Delta\nk(t).
\eqf
Thus, the perturbed mean anomaly is the sum of the now time-dependent mean anomaly at epoch $\eta(t)$ and a function of time $\varrho(t)$ whose derivative is equal to the (perturbed) mean motion, i.e.,
\begin{align}
\mathcal{M}^\textrm{pert}(t)\nonumber &=\eta(t) + \varrho(t)=\eta(t) + \int_{t_0}^{\,t}\,\nk^\textrm{pert}\ton{t^{'}}\,\mathrm{d}t^{'}=\\ \nonumber \\
\nonumber &= \eta+\Delta\eta(t) + \\ \nonumber \\
& + \nk\,\ton{t-t_0} + \int_{t_0}^{\,t}\,\Delta\nk\ton{t^{'}}\,\mathrm{d}t^{'}.
\end{align}
The resulting change $\Delta\mathcal{M}$ of the mean anomaly with respect to the unperturbed case is, thus,
\eqi
\Delta\mathcal{M}(t)= \mathcal{M}^\textrm{pert}(t)-\mathcal{M}(t)= \Delta\eta(t)+\Phi(t),
\eqf
where we defined
\eqi
\Phi(t) \doteq \int_{t_0}^{\,t}\,\Delta\nk\ton{t^{'}}\,\mathrm{d}t^{'}.\lb{Pphi}
\eqf
as a function whose derivative yields the perturbation of the mean motion.
In \rfr{Pphi}, the instantaneous shift of the mean motion due to the time-varying semimajor axis\footnote{It should be recalled that we kept $\mu$ constant.} $a$
is
\begin{align}
\Delta\nk(t) \nonumber & = -\rp{3}{2}\,\rp{\nk}{a}\,\Delta a(t)= \\ \nonumber \\
\nonumber &= -\rp{3}{2}\,\rp{\nk}{a}\,\int_{f_0}^{\,f}\,\dert{a}{t}\,\dert{t}{f^{'}}\,\mathrm{d}f^{'} = \\ \nonumber \\
& = -\rp{3}{2}\,\rp{\nk}{a}\,\Delta a\ton{f_0,\,f},\lb{Dn}
\end{align}
so that
\eqi
\Phi(t) = -\rp{3}{2}\,\rp{\nk}{a}\,\int_{f_0}^{\,f}\,\Delta a\ton{f_0,\,f^{'}}\,\dert{t}{f^{'}}\,\mathrm{d}f^{'} = \Phi\ton{f_0,\,f}.\lb{Phi}
\eqf
The shifts $\Delta a(t)$ and $\Delta\eta(t)$ can be perturbatively calculated by evaluating the right-hand-sides of the Gauss equations for their rates of change \cite{Bertotti03}
\begin{align}
\dert{a}{t} \lb{dadt} &= \rp{2}{\nk\,\sqrt{1-e^2}}\,\qua{e\,A_R\,\sin f + \ton{\rp{p}{r}}\,A_T},\\ \nonumber \\
\dert\eta{t} \lb{detadt} \nonumber &= -\rp{2}{\nk\,a}\,A_R\,\ton{\rp{r}{a}}-\\ \nonumber \\
& - \rp{\ton{1-e^2}}{\nk\,a\,e}\,\qua{-A_R\,\cos f +A_T\,\ton{1+\rp{r}{p}}\,\sin f},
\end{align}
onto the unperturbed Keplerian ellipse.
In \rfrs{dadt}{detadt}, $e$ is the eccentricity, $p\doteq a\ton{1-e^2}$ is the semilatus rectum, $r = p/\ton{1+e\cos f}$ is the (unperturbed) distance of the test particle from the primary, and $A_R$, $A_T$ are the projections of the perturbing pK acceleration $\bds A$ onto the radial and transverse directions, respectively.
The derivative of $t$ with respect to $f$ entering \rfrs{Dn}{Phi} is, up to terms of the first order in the perturbing acceleration $A$,
\eqi
\dert{t}{f} \simeq \rp{r^2}{\sqrt{\mu\,p}}+\mathcal{O}\ton{A}.
\eqf
Depending on the disturbing acceleration, $\Phi(t)$ is linear in time if the average over an orbital period $\Pb$ of its rate of change
\eqi
\ang{\overset{\cdot}{\Phi}(t)} =  -\rp{3}{4\,\uppi}\,\rp{\nk^2}{a}\,\int_{f_0}^{\,f_0+2\uppi}\,\Delta a\ton{f_0,\,f^{'}}\,\dert{t}{f^{'}}\,\mathrm{d}f^{'}
\eqf
is constant. Otherwise, it  may exhibit a more complex temporal pattern, as when the semimajor axis $a$ undergoes a secular change due to, e.g., some non-gravitational perturbing accelerations as in artificial satellites' dynamics. In general, the calculation of $\Phi(t)$ is rather cumbersome since it involves two integrations. Moreover, it depends on $f_0$.

From such considerations it follows that, at first sight, using the mean anomaly $\mathcal{M}(t)$ may not be a wise choice because of the disturbances introduced by $\Phi(t)$, especially in non-trivial scenarios in which several perturbing accelerations of different nature act simultaneously on the test particle inducing non-vanishing long-term  effects on the semimajor axis $a$.  Actually, we will show that it may not be the case in practical satellite data reductions if certain conditions are fulfilled.
%Even when it is not the case, as for the solar system's planets, the dependence of $\Phi(t)$ from $f_0$ is rather uncomfortable.
On the contrary, the mean anomaly at the epoch $\eta$, which is one of the six osculating Keplerian orbital elements in the perturbed restricted two-body problem, is not affected by such drawbacks. As such, it can be safely used, at least in principle, as an additional piece of information to improve some tests of pN gravity on the same foot of $\omega$ and $\Omega$. This fact seems to have gone unnoticed so far in the literature, as in the case of Ciufolini and Pavlis \cite{2005NewA...10..636C}.
\subsection{The mean longitude}
Similar considerations hold for the mean longitude $l$ defined as
\eqi
l(t)\doteq\varpi+\mathcal{M}(t),
\eqf
where
\eqi
\varpi\doteq\Omega+\omega
\eqf
is the longitude of pericentre. If a disturbing acceleration $\bds A$ is present, it can be expressed in terms of the mean longitude at epoch\footnote{It is more suited than $\eta$ at low orbital inclinations \cite{Bertotti03}.} $\epsilon$ \cite{Nobilibook87,Sof89,1991ercm.book.....B,Bertotti03} as
\begin{align}
l^\mathrm{pert}(t) \nonumber &=\epsilon(t) + \int_{t_0}^{\,t}\,\Delta\nk\ton{t^{'}}\,\mathrm{d}t^{'}=\\ \nonumber \\
&= \epsilon+\Delta\epsilon(t)+\int_{t_0}^{\,t}\,\Delta\nk\ton{t^{'}}\,\mathrm{d}t^{'},
\end{align}
so that its shift is
\eqi
\Delta l(t)=\Delta\epsilon(t)+\Phi(t).
\eqf
The shift $\Delta\epsilon(t)$ of the mean longitude at epoch can be worked out by means of the Gauss equation
for its variation \cite{Bertotti03}
\begin{align}
\dert\epsilon{t} \lb{depsdt} \nonumber &= -\rp{2}{\nk\,a}\,A_R\,\ton{\rp{r}{a}} + \rp{e^2}{1 + \sqrt{1-e^2}}\,\dert\varpi{t} + \\ \nonumber \\
& + 2\,\sqrt{1-e^2}\,\sin^2\ton{\rp{I}{2}}\,\dert{\Omega}{t},
\end{align}
where \cite{Bertotti03}
\begin{align}
\dert{\Omega}{t} \lb{dOgdt}& = \rp{1}{\nk\,a\,\sqrt{1-e^2}\,\sin I}\,A_N\,\ton{\rp{r}{a}}\,\sin u, \\ \nonumber \\
\dert{\varpi}{t} \lb{dodgdt} \nonumber & = \rp{\sqrt{1-e^2}}{\nk\,a\,e}\,\qua{-A_R\,\cos f + A_T\,\ton{1+\rp{r}{p}}\,\sin f }  +\\ \nonumber \\
&+  2\,\sin^2\ton{\rp{I}{2}}\,\dert{\Omega}{t}.
\end{align}
In \rfr{dOgdt}, $A_N$ is the projection of the perturbing acceleration $\bds A$ onto the out-of-plane direction, while
\eqi
u\doteq\omega + f
\eqf is the argument of latitude.
\section{The secular rates of change of $\eta(t),\,\epsilon(t),\,\Phi(t)$ for some pN accelerations}\lb{Greffects}
Here, we will preliminarily look at the effects due to the standard general relativistic pN accelerations induced by the static, gravitoelectric (Schwarzschild, Section\,\ref{Schw}) and stationary, gravitomagnetic (Lense-Thirring, Section\,\ref{Lets}) components of the spacetime of an isolated rotating body. We will not restrict to almost circular orbits; furthermore, we will allow the primary's spin axis $\bds{\hat{S}}$, entering the Lense-Thirring acceleration, to assume any orientation in space.
\subsection{The 1pN gravitoelectric Schwarzschild-like acceleration}\lb{Schw}
To the first pN order (1pN), the relative acceleration for two pointlike bodies of masses $m_\mathrm{A},\,m_\mathrm{B}$ separated by a distance $r$ and moving with relative velocity $\mathbf{v}$ is \cite{1985AIHS...43..107D,Sof89}
\begin{align}
{\bds A}^{\textrm{1pN}} \nonumber &= \frac{\mu_\textrm{tot}}{c^2\,r^2}\grf{\qua{\ton{4 + 2\,\zeta}\frac{\mu_\textrm{tot}}{r}-\right.\right.\\ \nonumber \\
\nonumber &\left.\left. - \ton{1 + 3\,\zeta}{\mathrm{v}}^2 +\frac{3}{2}\,\zeta\,{\mathrm{v}}_r^2}\,\bds{\hat{r}} + \right.\\ \nonumber \\
&\left. + \ton{4 - 2\,\zeta}\,{\mathrm{v}}_r\,\bds{\mathrm{v}}},\lb{megapN}
\end{align}
where $c$ is the speed of light in vacuum,
\eqi
\mu_\textrm{tot}=G\,\ton{m_\textrm{A} + m_\textrm{B}}
\eqf
is the total gravitational parameter of the binary system,
\eqi
{\mathrm{v}}_r \doteq \bds{\mathrm{v}}\bds\cdot\bds{\hat{r}}
\eqf
is the the radial velocity of the relative orbital motion,
and
\eqi
\zeta\doteq \rp{m_\mathrm{A}m_\mathrm{B}}{\ton{m_\mathrm{A} + m_\mathrm{B}}^2},\,0\leq\zeta\leq \rp{1}{4}.
\eqf
In Sections\,\ref{PhiSc}\,to\,\ref{etaepsSc}, we will work out the effect of \rfr{megapN} on $\Phi$, and $\eta$ and $\epsilon$, respectively.
\subsubsection{The shift $\Phi(t)$ due to the variation of the mean motion}
\begin{figure}[htb]
\centering
\centerline{
\vbox{
\begin{tabular}{c}
\epsfysize= 7.0 cm\epsfbox{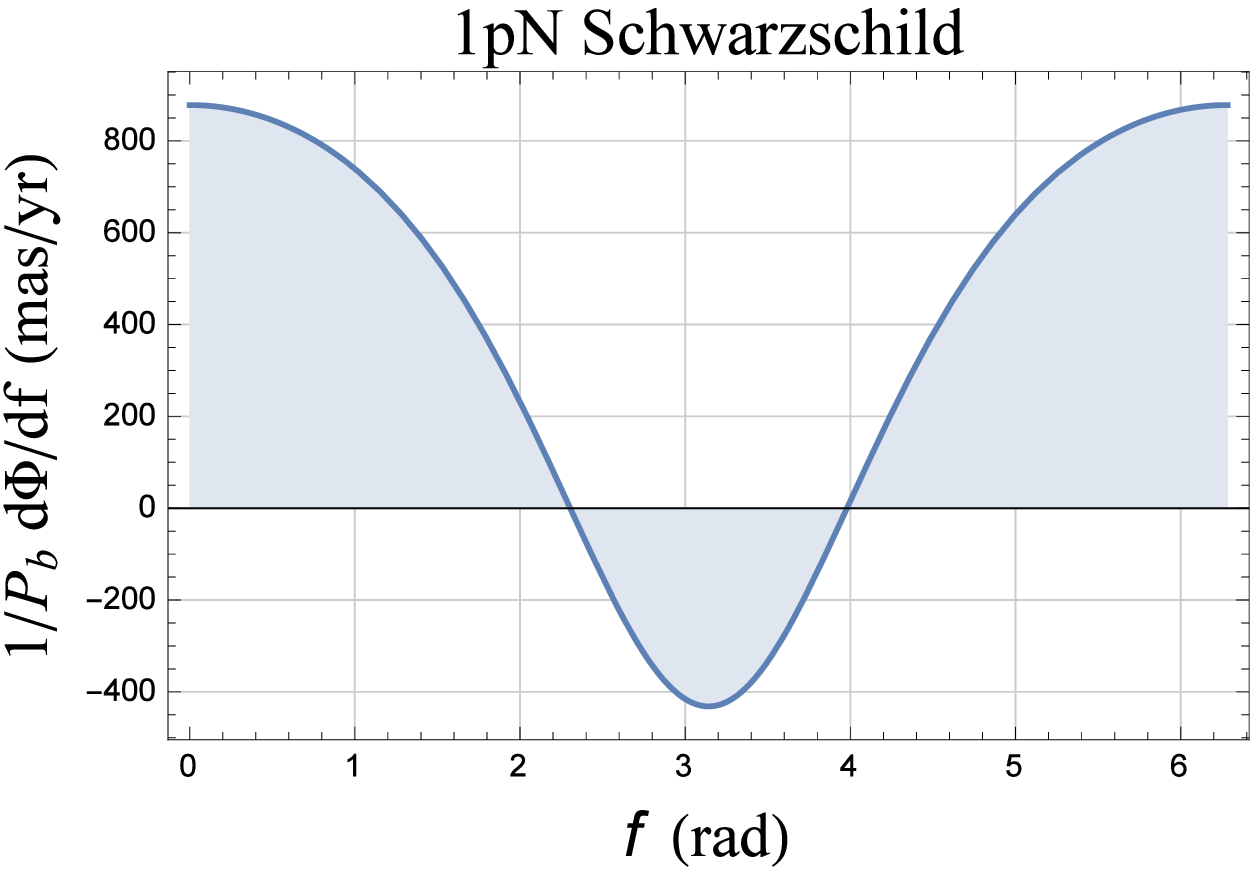}\\
\epsfysize= 7.0 cm\epsfbox{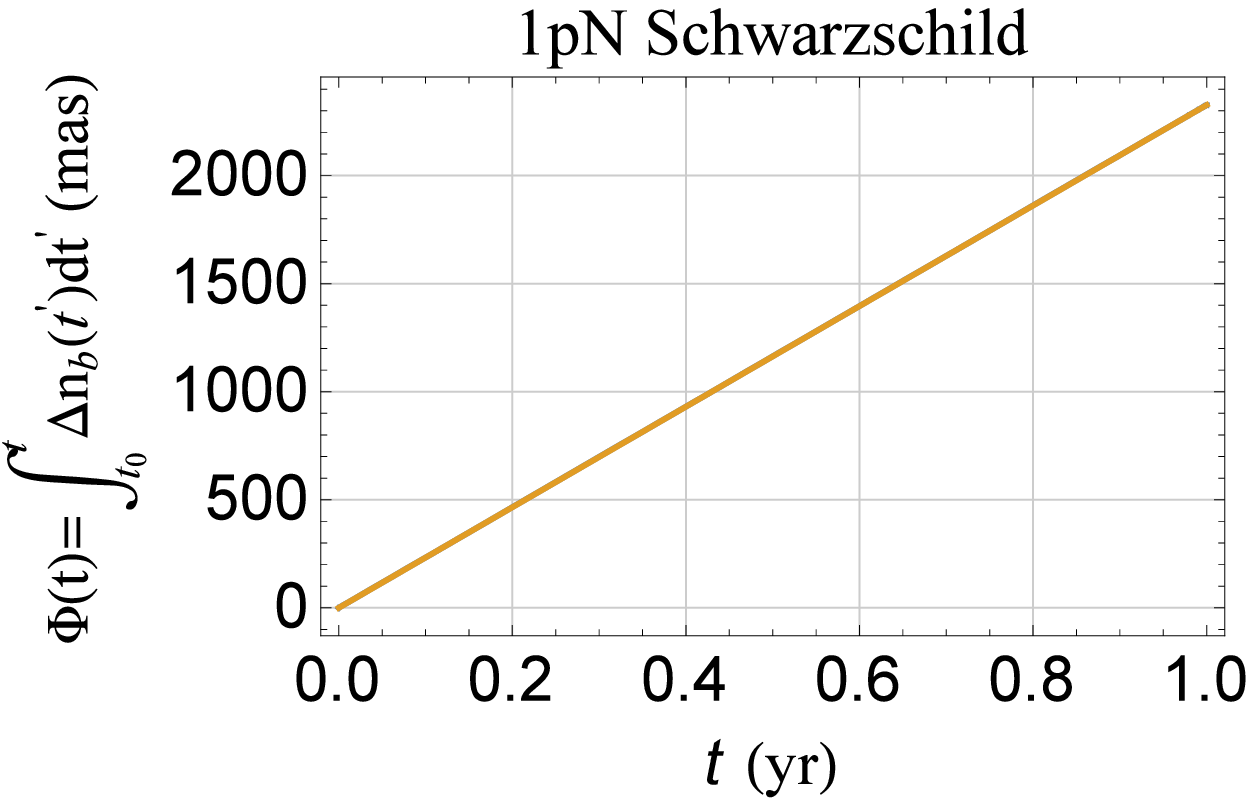}\\
\end{tabular}
}
}
\caption{Upper panel:
Plot of \rfr{inte}, computed for the orbital configuration of Table\,\ref{tavola1} and with $f_0 = 228\,\mathrm{deg}$, over a full orbital cycle of the true anomaly $f$. Its area, giving \rfr{dPhidt} in mas yr$^{-1}$,  amounts to $2,326.6\,\mathrm{mas\,yr}^{-1}$.
Lower panel: Numerically produced time series, in mas, of $\Phi(t)$ over 1 yr obtained by integrating the equations of motion in rectangular Cartesian coordinates for the fictitious Earth's satellite of Table\,\ref{tavola1}. The 1pN gravitoelectric Schwarschild-like acceleration was added to the Newtonian monopole. As initial value for the true anomaly, $f_0 = 228\,\mathrm{deg}$ was adopted. The slope of the linear trend amounts just to the area under the curve in the upper panel.  }\label{figura1}
\end{figure}

\lb{PhiSc}
By using \rfr{megapN} in \rfr{Dn} yields
\begin{align}
\Delta\nk\ton{f_0,\,f} \nonumber & = -\rp{3\,e\,\mu\,\nk\,\ton{\cos f-\cos f_0}}{4\,c^2\,a\,\ton{1-e^2}^2}\times\\ \nonumber \\
\nonumber &\times \grf{4 \qua{-7 + 3\,\zeta + e^2 \ton{-3 + 4\,\zeta}} +\right.\\ \nonumber \\
\nonumber &\left. + e \qua{e\,\zeta \cos 2f + 4 \ton{-5 + 4\,\zeta} \cos f_0 + \right.\right.\\ \nonumber \\
\nonumber &\left.\left. + 2 \cos f \ton{-10 + 8\,\zeta + e\,\zeta\,\cos f_0} + \right.\right.\\ \nonumber \\
&\left.\left. + e\,\zeta\,\cos 2f_0}}.
\end{align}
From it, the rate of change of $\Phi(t)$ averaged over one orbital period $\Pb$ can be straightforwardly worked out as
\begin{align}
\ang{\overset{\cdot}{\Phi}(t)} \nonumber &=\rp{1}{\Pb}\int_{t_0}^{\,t_0+\Pb}\,\dert{\Phi(t)}{t}\,\mathrm{d}t = \\ \nonumber \\
& = \rp{\nk}{2\uppi}\int_{f_0}^{\,f_0+2\uppi}\Delta\nk\ton{f_0,\,f}\,\dert{t}{f}\,\mathrm{d}f,\lb{dPhidt}
\end{align}
where
\begin{align}
\rp{\nk}{2\uppi}\,\Delta\nk\,\dert{t}{f} \nonumber \lb{inte}&=\rp{\nk}{2\uppi}\,\dert{\Phi}{t}\,\dert{t}{f} = \\ \nonumber \\
\nonumber & = \rp{1}{\Pb}\,\dert{\Phi}{f} = \\ \nonumber \\
\nonumber &= -\rp{3\,e\,\mu\,\nk\ton{\cos f -\cos f_0}}{8\,\uppi\,c^2\,a\,\sqrt{1 - e^2}\,\ton{1 +  e \cos f}^2}\times \\ \nonumber \\
\nonumber &\times \grf{4 \qua{-7 + 3\,\zeta + e^2 \ton{-3 + 4\,\zeta}} + \right.\\ \nonumber \\
\nonumber &\left. + e\,\qua{e\,\zeta \cos 2f + 4\,\ton{-5 + 4\,\zeta}\,\cos f_0 +\right.\right.\\ \nonumber \\
\nonumber &\left.\left. + 2\,\cos f\,\ton{-10 + 8\,\zeta + e\,\zeta \cos f_0} + \right.\right.\\ \nonumber \\
&\left.\left. + e\,\zeta \cos 2f_0}}.
\end{align}
From the analytical expression of the right-hand-side of \rfr{inte}, it turns out that the true anomaly $f$, and, thus, also the time $t$, appears only in trigonometric functions. This implies that, in this case, $\Phi(t)$ does not exhibit a polynomial temporal pattern, being, at most, linear in $t$ provided that \rfr{dPhidt} is not vanishing. Note also the dependence of \rfr{inte} on $f_0$.
We are not able to analytically calculate \rfr{dPhidt} unless a power expansion in $e$ of \rfr{inte} is made. Nonetheless, it is possible to perform a numerical integration of \rfr{dPhidt} for given values of the physical and orbital parameters entering it without any restriction on $e$. We successfully tested it for a fictitious cannonball geodetic satellite moving along an eccentric orbit, whose arbitrarily chosen physical and orbital parameters are displayed in Table\,\ref{tavola1},
\begin{table}[!htb]
\caption{Orbital and physical configuration of a fictitious terrestrial geodetic satellite. Since it is
$\rho_\textrm{LARES}=5.96\times 10^{-16}\,\textrm{kg\,m}^{-3}$ \cite{2017AcAau.140..469P}, and $\rho_\textrm{LAGEOS}=6.579\times 10^{-18}\,\textrm{kg\,m}^{-3}$ \cite{2015CQGra..32o5012L}, we, first, used them in   $\rho(h)=\rho_0\exp\qua{-\ton{h-h_0}\Lambda^{-1}}$, where $\rho_0$ and $h_0$ are, in general, referred to some reference height, to determine $\Lambda$ in the case $h_0=h_\textrm{LARES},\,h=h_\textrm{LAGEOS}$. Then, we used the so obtained characteristic length $\Lambda_\textrm{LR/L}=999.51\,\mathrm{km}$, valid in the range $h_\textrm{LARES}=1,442.06\,\textrm{km}<h<h_\textrm{LAGEOS}=5,891.96\,\textrm{km}$, to calculate $\rho_\textrm{max}$ for our orbital geometry. Instead, the value $\rho_\textrm{min}$ is just a guess which may be even conservative.
%The characteristic atmospheric length scale $L$ was inferred from $\rho(h)=\rho_0\exp\qua{-\ton{h-h_0}L^{-1}}$, where $h_0$ is some reference height, by %assuming $h_0=h_\textrm{min}=a(1-e)-R_\oplus,~h=h_\textrm{max}=a(1+e)-R_\oplus,~\rho_0=\rho_\textrm{max},~\rho(h)=\rho_\textrm{min}$. The same holds also for %the charged case. The values reported for the atmospheric densities, referred to the neutral hydrogen H component and to the charged part due to the ions %$\textrm{H}^{+},~\textrm{He}^{+},~\textrm{O}^{+}$, were retrieved from \cite{1990JGR....95.4881R} $(\rho_\textrm{max},~\rho^{+}_\textrm{max})$, while  %$\rho_\textrm{min},~\rho^{+}_\textrm{min}$ are just guesses.
The values of the satellite's physical parameters were taken from  \cite{2017AcAau.140..469P} ($m,~\Sigma,~C_\textrm{D}$).
}\lb{tavola1}
\begin{center}
\begin{tabular}{|l|l|l|}
\hline
Orbital and  physical parameter & Numerical value & Units\\
\hline
Mass (LARES) $m$ & $386.8$ & \textrm{kg}\\
Area-to-mass ratio $\Sigma$ (LARES) & $2.69\times 10^{-4}$ & $\textrm{m}^2~\textrm{kg}^{-1}$\\
%
%Electric charge $|q|$ (LAGEOS) & $1-100\times 10^{-11}$ & \textrm{C}\\
%
Neutral drag coefficient $C_\textrm{D}$ (LARES) & $3.5$ & - \\
%
%Charged drag coefficient $C_\textrm{D}^\textrm{ch}$ & $20$ & -\\
%
%Reflectivity coefficient $C_\textrm{R}$ (LARES) & $1.07$ & - \\
%
Semimajor axis $a$ & $12,500$ & \textrm{km}\\
Orbital period $P_\textrm{b}$ & $3.86$ & \textrm{hr}\\
Orbital eccentricity $e$ & $0.36$ & - \\
Perigee height $h_\textrm{min}$  & $1,621.86$ & \textrm{km}\\
Apogee height $h_\textrm{max}$  & $10,621.9$ & \textrm{km}\\
Orbital inclination $I$ & $63.43$ & \textrm{deg}\\
Argument of perigee $\omega$ & $0$ & \textrm{deg}\\
Period of the node $P_{\Omega}$ & $-1.76$ & \textrm{yr}\\
Period of the perigee $P_{\omega}$ & $-2,903.62$ & \textrm{yr}\\
Neutral atmospheric density at perigee  $\rho_\textrm{max}$  & $4.71\times 10^{-16}$ & \textrm{kg}~\textrm{m}$^{-3}$\\
Neutral atmospheric density at apogee $\rho_\textrm{min}$  & $1\times 10^{-20}$ & \textrm{kg}~\textrm{m}$^{-3}$\\
%
%Charged atmospheric density at perigee height  $\rho^{+}_\textrm{max}$ ($\textrm{H}^{+},~\textrm{He}^{+},~\textrm{O}^{+}$) & $2.9\times 10^{-17}$ & \textrm{kg}~\textrm{m}$^{-3}$\\
%
%Charged atmospheric density at apogee height  $\rho^{+}_\textrm{min}$ ($\textrm{H}^{+},~\textrm{He}^{+},~\textrm{O}^{+}$) & $1\times 10^{-19}$ & \textrm{kg}~\textrm{m}$^{-3}$\\
%
Characteristic atmospheric length scale  $\Lambda$  & $836.34$ & \textrm{km}\\
%
%Characteristic atmospheric length scale  $\lambda{+}$ ($\textrm{H}^{+},~\textrm{He}^{+},~\textrm{O}^{+}$) & $3,127.05$ & \textrm{km}\\
%
\hline
\end{tabular}
\end{center}
\end{table}
by numerically integrating its equations of motion in rectangular Cartesian coordinates, and by numerically performing the integral of \rfr{dPhidt} with \rfr{inte}.
Fig.\,\ref{figura1} displays the plot of \rfr{inte}, in milliarcseconds per year $\ton{\mathrm{mas\,yr}^{-1}}$, for the orbital parameters of Table\,\ref{tavola1}, and the numerically produced time series of $\Phi(t)$, in mas,  over 1 yr for the same orbital configuration; the agreement between the slope of $\Phi(t)$ and the area under the curve of \rfr{inte} is remarkable. From Fig.\,\ref{figura1}, it can be noted that, as expected, the 1pN Schwarzschild-like acceleration induces a secular variation on $\Phi(t)$ which has to be added to those affecting $\eta$ and $\epsilon$ displayed in Section\,\ref{etaepsSc}.
\clearpage{}

The opportunity offered by the exact expression of \rfr{inte} to calculate $\ang{\overset{\cdot}{\Phi}(t)}$ as per \rfr{dPhidt} is important also in astronomical and astrophysical scenarios, like the almost circular orbital motions of the major bodies of our solar system and the much more eccentric ones of various types of binary systems (extrasolar planets, binary stars, binary pulsars hosting at least one emitting neutron star, stellar systems revolving around supermassive galactic black holes, etc.), in which secular variations of the semimajor axis $a$-or even of the masses involved-are absent or negligible with respect to either the duration of the typical data analyses or to the observational accuracy. Indeed, in all such cases, the perturbed evolution of the mean anomaly can, in principle, be monitored as well, and $\ang{\overset{\cdot}{\Phi}(t)}$ may represent an important contribution to the overall long-term rate of change of  $\mathcal{M}(t)$. Suffice it to say that, in the case of Mercury and the Sun, it is
\eqi
\ang{\overset{\cdot}{\Phi}} = 210.3\,\mathrm{arcsec\,cty}^{-1}.
\eqf
\subsubsection{The mean anomaly at epoch $\eta$ and the mean longitude at epoch $\epsilon$}\lb{etaepsSc}
The Gauss equations for the variation of $\eta$ and $\epsilon$ (\rfr{detadt} and \rfr{depsdt}) allow to straightforwardly work out their secular rates of change which turn out to be
\begin{align}
\ang{\overset{\cdot}{\eta}} \lb{etpN}& = \rp{\mu\,\nk\,\qua{-15 + 6\sqrt{1-e^2} + \ton{9 - 7\sqrt{1-e^2}}\,\zeta}}{c^2\,a\,\sqrt{1-e^2}}, \\ \nonumber \\
\ang{\overset{\cdot}{\epsilon}} \lb{epspN}& =-\rp{\mu\,\nk\,\qua{-9+15\sqrt{1-e^2} +e^2\ton{6-7\,\zeta} + \ton{7-9\sqrt{1-e^2}}\,\zeta  }}{c^2\,a\,\ton{1-e^2}}.
\end{align}
They were confirmed by a numerical integration of the equations of motion in the case of the satellite's orbital configuration of Table\,\ref{tavola1} which returned linear times series whose slopes agree with \rfrs{etpN}{epspN}. In the case of Mercury and the Sun, \rfrs{etpN}{epspN} yield
\begin{align}
\ang{\overset{\cdot}{\eta}} & = -127.986\,\mathrm{arcsec\,cty}^{-1}, \\ \nonumber \\
\ang{\overset{\cdot}{\epsilon}} & = -85.004\,\mathrm{arcsec\,cty}^{-1}.
\end{align}
\subsection{The 1pN gravitomagnetic Lense-Thirring acceleration}\lb{Lets}
In the case of the 1pN gravitomagnetic Lense-Thirring acceleration \cite{Sof89} induced by the spin dipole moment of the central mass, i.e. its proper angular momentum $\bds S$,  on a test particle orbiting it with velocity $\mathbf{v}$
\eqi
{\bds A}^{\mathrm{LT}} \lb{1pNS}  = \frac{2\,G\,S}{c^2\,r^3}\qua{3\,\xi\,\bds{\hat{r}}\bds\times\bds{\mathrm{v}} + \bds{\mathrm{v}}\bds\times\kap}
,
\eqf  it turns out that
\eqi
\Delta\nk\ton{f_0,\,f} = 0
\eqf
for an arbitrary orientation of the body's spin axis $\bds{\hat{S}}$ in space. Thus, it is
\eqi
\Phi(t)=0.\lb{cazzo}
\eqf
It implies that the claims by Ciufolini and Pavlis \cite{2005NewA...10..636C} about an alleged non-vanishing perturbing effect of the gravitomagnetic field of the Earth on both the semimajor axis $a$ and the mean motion $\nk$ of a  satellite are, in fact, erroneous for \textit{any} spacecraft.

Moreover, it is also
\begin{align}
\ang{\overset{\cdot}{\eta}} &=0, \\ \nonumber \\
\ang{\overset{\cdot}{\epsilon}} \lb{epsLT}&= \rp{2G\bds S\bds\cdot\qua{-2\bds{\hat{h}}+\ton{\csc I-\cot I}\bds{\hat{m}} } }{c^2 a^3\ton{1-e^2}^{3/2}}.
\end{align}
for any $\bds{\hat{S}}$ as well. In \rfr{epsLT},
\eqi
\bds{\hat{h}}=\grf{\sin I\sin\Omega,\,-\sin I\cos\Omega,\,\cos I}
\eqf
is the unit vector directed along the orbital angular momentum along the out-of-plane direction, while
\eqi
\bds{\hat{m}}=\grf{-\cos I\sin\Omega,\,\cos I\cos\Omega,\,\sin I}
\eqf
is the unit vector directed transversely to the line of the nodes in the orbital
plane. In the case of an Earth's satellite, by assuming, as usual, an equatorial coordinate system with its reference $z$ axis directed along $\bds{\hat{S}}$, \rfr{epsLT} reduces to
\eqi
\ang{\overset{\cdot}{\epsilon}} = \rp{2GS\ton{1-3\cos I}}{c^2 a^3\ton{1-e^2}^{3/2}}.\lb{inculata}
\eqf
\rfr{cazzo} and \rfr{inculata} show that the claim by Ciufolini and Pavlis \cite{2005NewA...10..636C} \virg{[\ldots] let us, for example, consider a satellite at the LAGEOS altitude, the Lense-Thirring effect on its mean longitude is of the order of 2 m/y, [\ldots]} is wrong. Indeed, the gravitomagnetic linear shift corresponding to \rfr{inculata} amounts to $3.68\,\mathrm{m\,yr}^{-1}$ for LAGEOS; it is an enormous discrepancy with respect to the statement by Ciufolini and Pavlis \cite{2005NewA...10..636C} since the present-day accuracy in reconstructing the orbits of the laser-ranged satellites of the LAGEOS type is notoriously at the $\simeq 1-0.5\,\mathrm{cm}$ level.
\section{The secular rates of change of $\eta(t),\,\epsilon(t),\,\Phi(t)$ for some Newtonian perturbing accelerations}\lb{Oblo}
Here, we will deal with the impact of the oblateness of the primary (Section\,\ref{JEI2}), whose spin axis $\bds{\hat{S}}$ is assumed arbitrarily oriented in space, and of the atmospheric drag (Section\,\ref{Drag}). The small eccentricity approximation for the satellite's orbit will not be adopted. Such classical accelerations represent two of the most important sources of systematic errors in accurate tests of pN gravity with artificial satellites. On the other hand, they can be considered interesting in themselves if one is interested in better characterizing the shape and the inner mass distribution of the primary like, e.g., a star, at hand, and the properties of the atmosphere of the orbited planet.
\subsection{The quadrupole mass moment $J_2$}\lb{JEI2}
To the Newtonian level, the external potential of an oblate body at the outside position $\bds r$ is
\eqi
U\ton{\bds r} = U_0 + \Delta U_2 = -\frac{\mu}{r}\qua{1-\ton{\frac{R_\textrm{e}}{r}}^2\,J_2\,\mathcal{P}_2\ton{\xi}},
\eqf
where  $J_2$ is the first even zonal harmonic coefficient of the multipolar expansion of its classical gravitational potential,
\eqi
\xi\doteq \bds{\hat{S}}\bds\cdot\bds{\hat{r}}
\eqf
is the cosine of the angle between the primary's spin axis and the particle's position,
and
\eqi
\mathcal{P}_2\ton{\xi} = \rp{3\xi^2 - 1}{2}
\eqf
is the Legendre polynomial of degree 2.
The Newtonian acceleration due to $J_2$ experienced by a test particle orbiting the distorted axisymmetric primary is
\begin{align}
{\bds A}^{\textrm{N}J_2} \nonumber & = -\bds\nabla \Delta U_{J_2}= \\ \nonumber \\
& = \frac{3\,\mu\,R_\textrm{e}^2\,J_2}{2\,r^4}\qua{\ton{5\,\xi^2 - 1}\,\bds{\hat{r}} - 2\,\xi\,\kap}.\lb{NJ2}
\end{align}
In Sections\,\ref{PhiJ2}\,to\,\ref{etaepsJ2}, we will work out its impact on $\Phi(t)$, and $\eta$ and $\epsilon$, respectively.
\subsubsection{The shift $\Phi(t)$ due to the variation of the mean motion}
\begin{figure}[htb]
\centering
\centerline{
\vbox{
\begin{tabular}{c}
\epsfysize= 7.0 cm\epsfbox{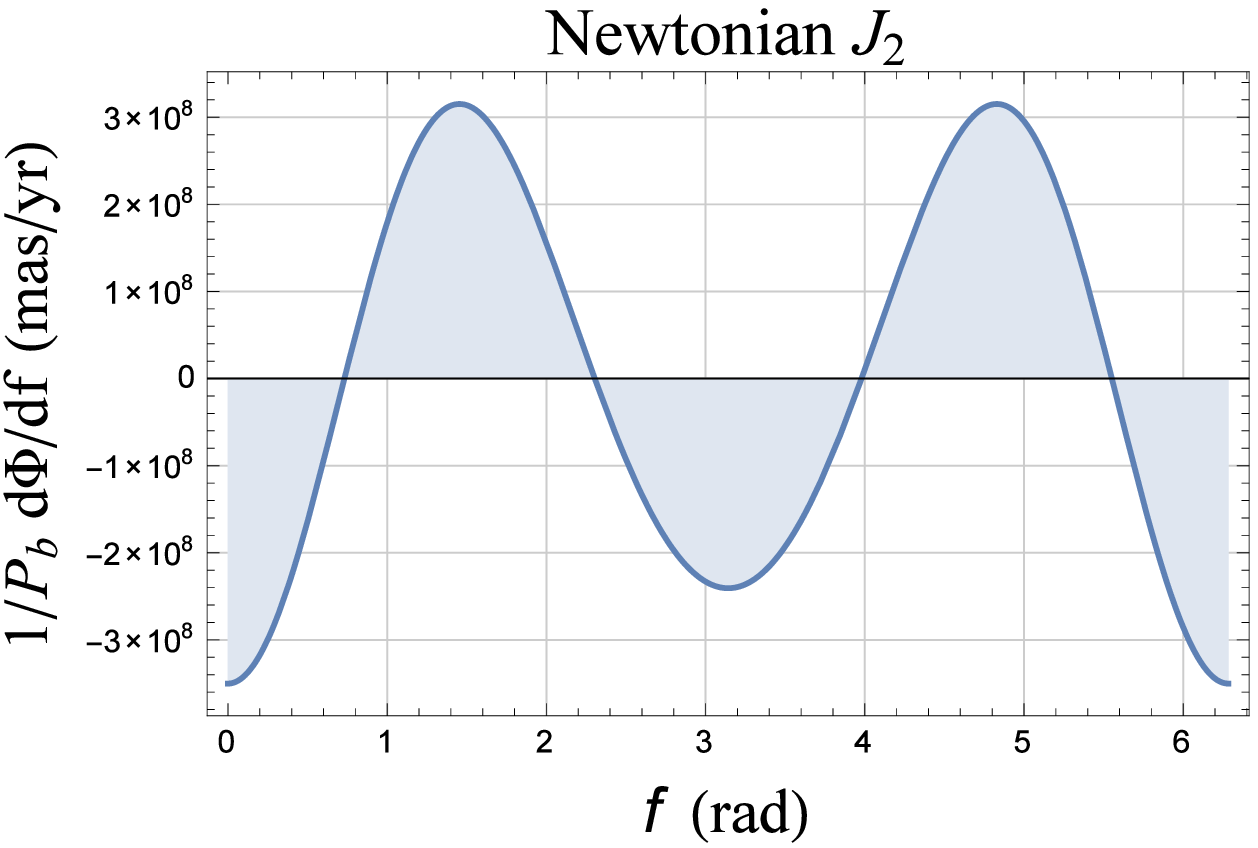}\\
\epsfysize= 7.0 cm\epsfbox{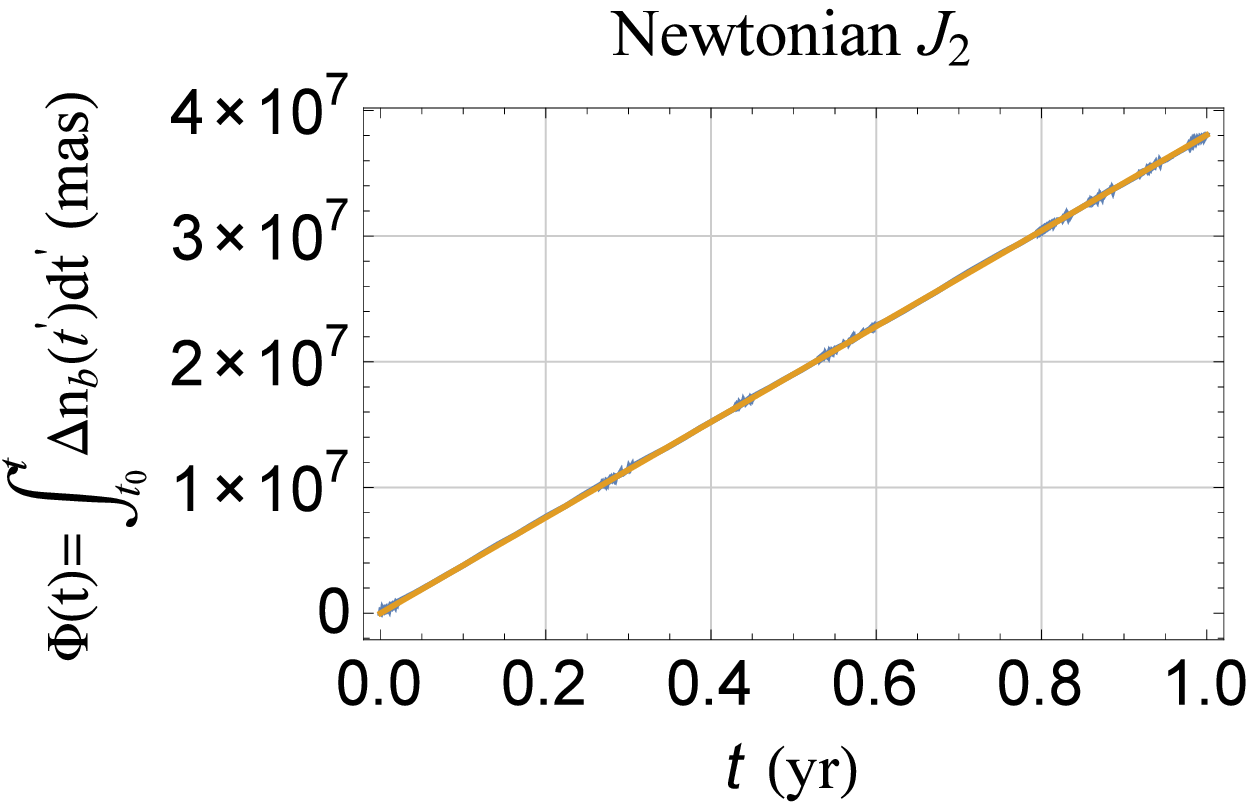}\\
\end{tabular}
}
}
\caption{Upper panel:
Plot of \rfr{scassa}, computed for the orbital configuration of Table\,\ref{tavola1} and with $f_0 = 228\,\mathrm{deg}$, over a full orbital cycle of the true anomaly $f$. Its area, giving $\ang{\overset{\cdot}{\Phi}}$ in mas yr$^{-1}$, turns out to be equal to  $3.8\times 10^7\,\mathrm{mas\,yr}^{-1}$.
Lower panel: Numerically produced time series, in mas, of $\Phi(t)$ over 1 yr obtained by integrating the equations of motion in rectangular Cartesian coordinates for the fictitious Earth's satellite of Table\,\ref{tavola1}. The Newtonian acceleration of \rfr{NJ2} due to $J_2$  was added to the Newtonian monopole. As initial value for the true anomaly, $f_0 = 228\,\mathrm{deg}$ was adopted. The  slope of the linear trend amounts just to the area under the curve in the upper panel.  }\label{figura5}
\end{figure}

\lb{PhiJ2}
It turns out that
\eqi
\ang{\overset{\cdot}{\Phi}}\neq 0,
\eqf
so that $\Phi(t)$, which depends on $f_0$, is linear in time. It is not possible to explicitly display the analytical expression which we obtained for $\ton{1/\Pb}\,\mathrm{d}\Phi/\mathrm{d}f$ in the case of an arbitrary orientation of $\bds{\hat{S}}$ in space because of its cumbersomeness. However, it can be fruitfully used with, e.g., any astronomical binary systems  since, in general, their spin axes are not aligned with the line of sight which, usually, is assumed as reference $z$ axis of the coordinate systems adopted.  In regard to an Earth's satellite, whose motion is customarily studied in an equatorial coordinate system whose reference $z$ axis is aligned with $\bds{\hat{S}}$, we have
\begin{align}
\rp{\nk}{2\uppi}\,\Delta\nk\,\dert{t}{f} \nonumber & =\rp{\nk}{2\uppi}\,\dert{\Phi}{t}\,\dert{t}{f}=\\ \nonumber \\
\nonumber & = \rp{1}{\Pb}\,\dert{\Phi}{f} = \\ \nonumber \\
&- \rp{3\,\nk\,R_\mathrm{e}^2\,J_2}{64\,\uppi\,a^2\,\ton{1-e^2}^{3/2}\,\ton{1+e\cos f}^2}\mathcal{J},\lb{scassa}
\end{align}
with
\begin{align}
\mathcal{J}\nonumber &= -e\,\qua{12\,\cos f_0  + e\,\ton{-6\,\cos 2 f  - e\,\cos 3 f  + \right.\right. \\ \nonumber \\
\nonumber &\left.\left. + 4\,e\,\cos^3 f_0 + 6\,\cos 2 f_0 }}\,\ton{1 + 3\,\cos 2 I} - \\ \nonumber \\
\nonumber & -3 \ton{-4 \ton{2 + 3\,e^2}\,\cos 2 f  + 8\,\cos 2 f_0  + \right.\\ \nonumber \\
\nonumber &\left. + e\,\grf{12\,\ton{\cos f_0  + \cos 3 f_0 } + \right.\right.\\ \nonumber \\
\nonumber & +\left.\left. e\,\qua{-6\,\cos 4 f  + 4\,\ton{3 + 2\,e\,\cos^3 f_0}\,\cos 2 f_0  + \right.\right.\right.\\ \nonumber \\
\nonumber &\left.\left.\left. + 6\,\cos 4 f_0 }}}\,\sin^2 I\,\cos 2\omega + \\ \nonumber \\
\nonumber & + 24\,e^3\,\cos^3 f\,\cos 2  u\,\sin^2 I + \\ \nonumber \\
\nonumber & + 3\,\cos f \,\qua{e\,\ton{4 + e^2}\,\ton{1 + 3\,\cos 2 I} + \right.\\ \nonumber \\
\nonumber &\left. + 24\,e\,\sin^2 I\,\cos 2u} + \\ \nonumber \\
\nonumber & + 3\,\qua{-4\,\ton{2 + 3\,e^2 + 3\,e^2\,\cos 2 f}\,\sin 2 f  +\right. \\ \nonumber \\
&\left. +  8\,\ton{1 + e\,\cos f_0 }^3\,\sin 2 f_0 }\,\sin^2 I\,\sin 2 \omega.
\end{align}
The numerical value of the area under the plot of \rfr{scassa}, depicted in the upper panel of Fig.\,\ref{figura5},  is confirmed by the time series for $\Phi(t)$ produced by numerically integrating the equations of motion of the fictitious satellite of Table\,\ref{tavola1}, and displayed in the lower panel of Fig.\,\ref{figura5}.
\clearpage{}
\subsubsection{The mean anomaly at epoch $\eta$ and the mean longitude at epoch $\epsilon$}\lb{etaepsJ2}
The Gauss equations
for the variations of $\eta$ and $\epsilon$ (\rfr{detadt} and \rfr{depsdt}) allow to straightforwardly obtain
\begin{align}
\ang{\overset{\cdot}{\eta}} \lb{etaJ2}& = \rp{3\,\nk\,R_\mathrm{e}^2\,J_2\grf{2-3\qua{\ton{\bds{\hat{S}}\bds\cdot\bds{\hat{l}}}^2+\ton{\bds{\hat{S}}\bds\cdot\bds{\hat{m}}}^2} }}{4\,a^2\,\ton{1-e^2}^{3/2}}, \\ \nonumber \\
\ang{\overset{\cdot}{\epsilon}} \lb{epsJ2}\nonumber & = \rp{3\,\nk\,R_\mathrm{e}^2\,J_2}{4\,a^2}
\grf{
\rp{2-3\qua{\ton{\bds{\hat{S}}\bds\cdot\bds{\hat{l}}}^2+\ton{\bds{\hat{S}}\bds\cdot\bds{\hat{m}}}^2}}{\ton{1-e^2}^{3/2}}+\right.\\ \nonumber \\
&\left.+ \rp{2-3\qua{\ton{\bds{\hat{S}}\bds\cdot\bds{\hat{l}}}^2+\ton{\bds{\hat{S}}\bds\cdot\bds{\hat{m}}}^2} -2\ton{\bds{\hat{S}}\bds\cdot\bds{\hat{h}}}\ton{\bds{\hat{S}}\bds\cdot\bds{\hat{m}}}\ton{1-\cot I}}{\ton{1-e^2}^2}
},
\end{align}
where
$\bds{\hat{l}}=\grf{\cos\Omega,\,\sin\Omega,\,0}$ is the unit vector directed along the line of the nodes such that $\bds{\hat{l}}\bds\cross\bds{\hat{m}}=\bds{\hat{h}}$.
Also \rfrs{etaJ2}{epsJ2} can be used with any astronomical binary system in view of their generality.
In the case of a coordinate system with its reference $z$ axis aligned with the body's spin axis, as in the case of an Earth's satellite referred to an equatorial coordinate system, \rfrs{etaJ2}{epsJ2} reduce to
\begin{align}
\ang{\overset{\cdot}{\eta}} & = \rp{3\,\nk\,R_\mathrm{e}^2\,J_2\ton{1 + 3\cos 2I}}{8\,a^2\,\ton{1-e^2}^{3/2}}, \\ \nonumber \\
\ang{\overset{\cdot}{\epsilon}} & = \rp{3\,\nk\,R_\mathrm{e}^2\,J_2\,\qua{3 + \sqrt{1 - e^2} - 4 \cos I + \ton{5 + 3 \sqrt{1 - e^2}} \cos 2I}}{8\,a^2\,\ton{1 - e^2}^2}.
\end{align}
%
%
%
%
%A more effective computational approach, especially as far as the zonal harmonics $J_\ell$ of degree $\ell$ higher than $2$ are concerned, consists of using %the Lagrange equations \cite{Bertotti03}
%\begin{align}
%\ang{\overset{\cdot}{\eta}} \lb{Lagreta}& = \rp{2}{\nk\,a}\derp{\ang{\Delta U_\ell}}{a} +\rp{\ton{1-e^2}}{\nk\,a^2\,e}\derp{\ang{\Delta U_\ell}}{e}, \\ %\nonumber \\
%
%\ang{\overset{\cdot}{\epsilon}} & = \rp{2}{\nk\,a}\derp{\ang{\Delta U_\ell}}{a}-\rp{\sqrt{1-e^2}\ton{1-\sqrt{1-e^2}}}{\nk\,a^2\,e}\derp{\ang{\Delta %U_\ell}}{e}-\rp{\tan\ton{I/2}}{\nk\,a^2\,\sqrt{1-e^2}}\derp{\ang{\Delta U_\ell}}{I}.
%\end{align}
%In them,
%\eqi
%\Delta U_\ell = \rp{\mu}{r}\,\ton{\rp{R_\mathrm{e}}{r}}^\ell\,J_\ell\,\mathcal{P}_\ell\ton{\xi},\,\ell=2,\,3,\,4\,\ldots
%\eqf
%is the zonal perturbing potential of degree $\ell$,
%where $\mathcal{P}_\ell\ton{\xi}$ is the Legendre polynomial of degree $\ell$.
%
\subsection{The atmospheric drag}\lb{Drag}
The atmospheric drag induces, among other things, a secular decrease of the semimajor axis $a$ which, in turn, has an impact on $\nk(t)$ and $\Phi(t)$.

For a  cannonball geodetic satellite, the drag acceleration can be expressed as
\eqi
{\bds A}_\textrm{D} = -\rp{1}{2}\,C_\textrm{D}\,\Sigma\,\rho\,V\,\bds V.\lb{adrag}
\eqf
In \rfr{adrag}, $C_\textrm{D},~\Sigma,~\rho,~\bds V$ are the dimensionless drag coefficient of the satellite, its area-to-mass ratio, the atmospheric density at its height, and its velocity with respect to the atmosphere, respectively. In the following, we will assume that the atmosphere co-rotates with the Earth. Thus, $\bds V$ is
\eqi
\bds V = \bds{\textrm{v}}-\bds\Psi\cross\bds r,\lb{roto}
\eqf
where $\bds\Psi$ is the Earth's angular velocity.
%In fact, a decrease of the co-rotation with the height is expected. \cite{2010EJPh...31.1013M} modeled it in two scenarios involving  a constant and %non-constant viscosity. In the first case, the second term in \rfr{roto} must be rescaled by $\ton{R_\textrm{e}/r}^3$, where $R_\textrm{e}$ is the Earth's %equatorial radius.
We will model the atmospheric density as
\eqi
\rho(r) = \rho_0\exp\qua{-\rp{\ton{r-r_0}}{\Lambda}},
\eqf
where $\rho_0$ refers to some reference distance $r_0$, while $\Lambda$ is the characteristic scale length. By assuming
\eqi
r_0=r_\textrm{min}=a\left(1-e\right),
\eqf
$\Lambda$ can be determined as
\eqi
\Lambda = -\rp{2\,a\,e}{\ln\ton{\rp{\rho_\textrm{min}}{\rho_\textrm{max}}}},
\eqf
where
\begin{align}
\rho_\textrm{min}&=\rho(r_\textrm{max}),\\ \nonumber \\
\rho_\textrm{max} & =\rho(r_\textrm{min})
\end{align}
are the values of the atmospheric density at the apogee and perigee heights, respectively.
Table~\ref{tavola1} shows the neutral  atmospheric density at the perigee height chosen as inferred from existing data on LAGEOS and LARES. On the other hand, the values reported for the apogee are purely speculative and should be regarded as subjected to huge uncertainties. Actually, even the density at a given height may not be regarded as truly constant because of a variety of geophysical phenomena characterized by quite different time scales. Anyway, in order to have an order-of-magnitude evaluation of the perturbing action of \rfr{adrag} on the motion of the fictitious satellite of Table\,\ref{tavola1}, we will make our calculation  by keeping $\rho_0$ fixed during one orbital period $P_\textrm{b}$. An exact analytical calculation without recurring to any approximation in both $e$ and $\nu\doteq \Psi/\nk$ is difficult.

In Sections\,\ref{dragPhi}\,to\,\ref{drageteps}, we will calculate the impact of \rfr{adrag} on $\Phi(t)$, and $\eta$ and $\epsilon$, respectively.
\subsubsection{The shift $\Phi(t)$ due to the variation of the mean motion}
\begin{figure}[htb]
\centering
\centerline{
\vbox{
\begin{tabular}{c}
\epsfysize= 7.0 cm\epsfbox{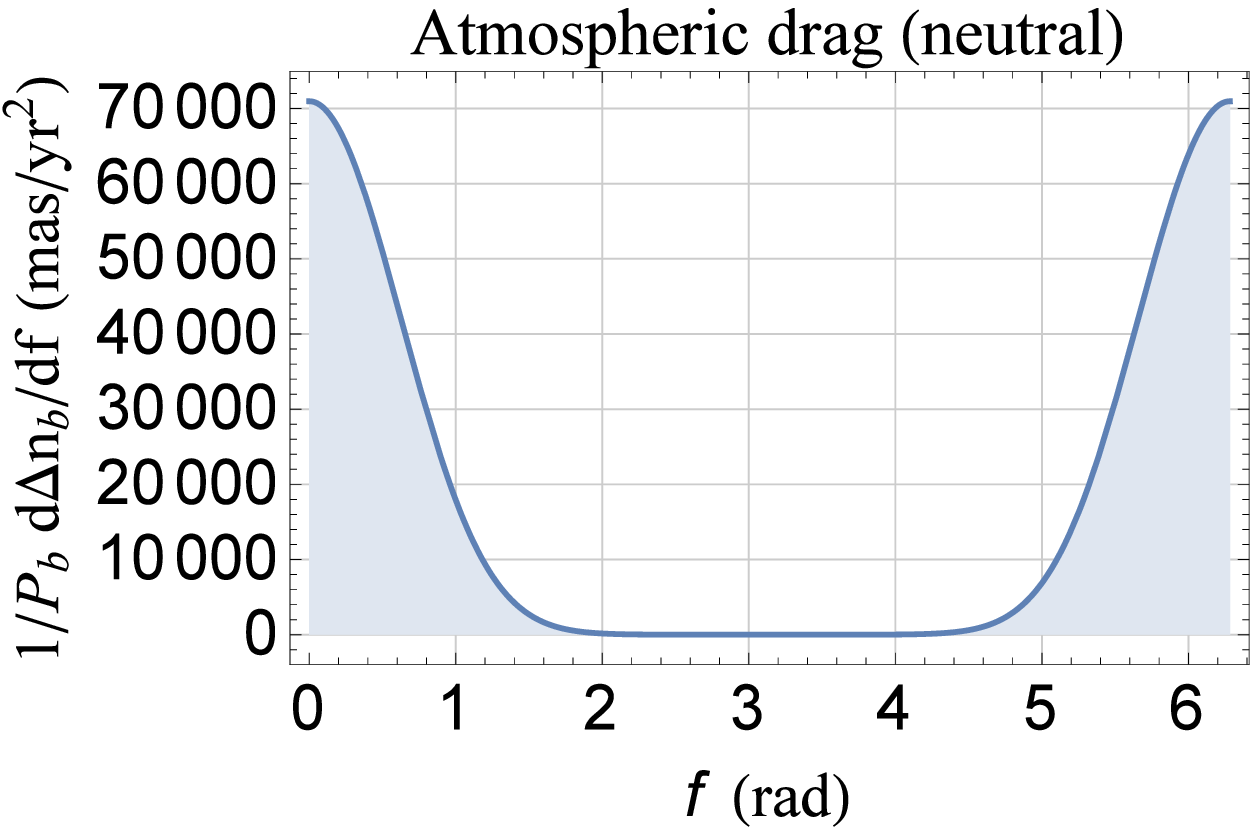}\\
\epsfysize= 7.0 cm\epsfbox{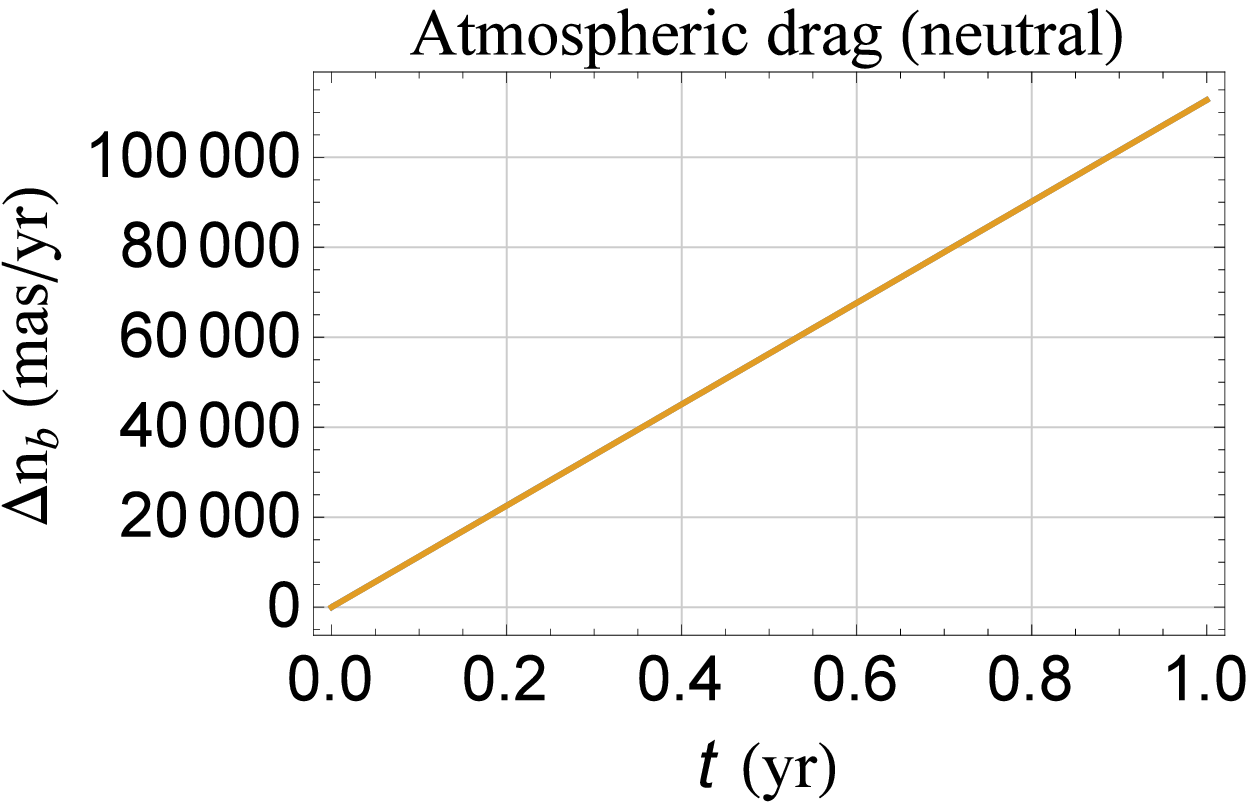}\\
\end{tabular}
}
}
\caption{Upper panel:
Plot of \rfr{uno}, computed for the orbital configuration of Table\,\ref{tavola1} and with $f_0 = 228\,\mathrm{deg}$, over a full orbital cycle of the true anomaly $f$. Its area, giving $\ang{\overset{.}{\Delta\nk}}$ in mas yr$^{-2}$, turns out to be equal to  $107,217\,\mathrm{mas\,yr}^{-2}$.
Lower panel: Numerically produced time series, in mas yr$^{-1}$, of $\Delta\nk(t)$ over 1 yr obtained by integrating the equations of motion in rectangular Cartesian coordinates for the fictitious Earth's satellite of Table\,\ref{tavola1}. The drag acceleration of \rfr{adrag} was added to the Newtonian monopole. As initial value for the true anomaly, $f_0 = 228\,\mathrm{deg}$ was adopted. The linear trend is apparent, and its slope amounts just to the area under the curve in the upper panel.  }\label{figura2}
\end{figure}

\begin{figure}[htb]
\centering
\centerline{
\vbox{
\begin{tabular}{c}
\epsfysize= 7.0 cm\epsfbox{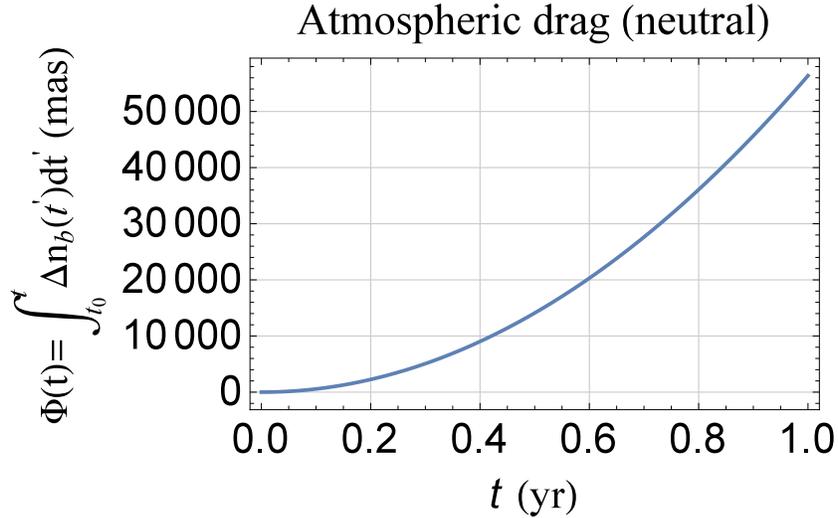}\\
\end{tabular}
}
}
\caption{
Numerically produced time series, in mas, of $\Phi(t)$ over 1 yr obtained by integrating the equations of motion in rectangular Cartesian coordinates for the fictitious Earth's satellite of Table\,\ref{tavola1}. The drag acceleration of \rfr{adrag} was added to the Newtonian monopole. As initial value for the true anomaly, $f_0 = 228\,\mathrm{deg}$ was adopted. The quadratic signature is apparent, and its final value is in agreement with what expected from Fig.\,\ref{figura2}.}\label{figura3}
\end{figure}

\lb{dragPhi}
Let us, now, start to look at $\Delta\nk(t)$ by means of \rfr{Dn}. We will show that it is linear in time because $\ang{{\overset{.}{\Delta\nk}}}\neq 0$.
The analytical expression of $1/\Pb\,\mathrm{d}\Delta\nk/\mathrm{d}f$ is
\begin{align}
\ton{\rp{\nk}{2\uppi}}\,\ton{-\rp{3}{2}\,\rp{\nk}{a}\,\dert{a}{f}}  \nonumber \lb{uno}&= \\ \nonumber \\
\nonumber & = \rp{3\,C_\textrm{D}\,\Sigma\,\rho\ton{f}\,\nk^2\,\sqrt{1-e^2}\,\mathcal{V}\ton{f}}{ 4\,\uppi\,\ton{1 + e\,\cos f}^2 }\times \\ \nonumber \\
\nonumber &\times \qua{1 + 2\,e\,\cos f + e^2 - \right.\\ \nonumber \\
&\left. - \nu\,\ton{1-e^2}^{3/2}\,\cos I},
\end{align}
where
\begin{align}
{\mathcal{V}}^2\ton{f} \lb{Vu}\nonumber & =1 - \nu\rp{2\,\ton{1 - e^2}^{3/2}\,\cos I}{1 + e^2 + 2\,e\,\cos f} + \\ \nonumber \\
&+ \nu^2\,\rp{\ton{1-e^2}^3 \ton{3 + \cos 2 I +2\,\sin^2 I\,\cos 2u } }{4\,\ton{1 + e\,\cos f}^2\,\ton{1 + e^2 + 2\,e\,\cos f} }.
\end{align}
%
%
%It is worthwhile noticing that, in general,
%\eqi
%\left|{\mathcal{V}}^2-1\right|\nless 1,
%\eqf
%being even possible that
%\eqi
%\left|{\mathcal{V}}^2-1\right|\gtrsim 1
%\eqf  for some values of $f$, thus preventing from expanding it in powers of $\nu$.
Since it is not possible to analytically integrate \rfr{uno} with \rfr{Vu} in the most general case without recurring to approximations in $e$ and $\nu$, we will plot it as a function of $f$ over a full orbital cycle and integrate it numerically for the physical and orbital parameters of Table\,\ref{tavola1}. The upper panel of Fig.\,\ref{figura2} depicts \rfr{uno}, while the lower panel displays the time series for $\Delta\nk(t)$ calculated  from a numerical integration of the satellite's equations of motion in rectangular Cartesian coordinates over 1 yr.
\clearpage{}
The fact that $\ang{\overset{.}{\Delta\nk}}\neq 0$ implies that $\Delta\nk$ is linear\footnote{Strictly speaking, it is, in general, true only for fast satellites orbiting in much less than a day, so that the term proportional to $\nu^2$ in \rfr{Vu}, which contains $\omega$, can be neglected. However, in the particular case of the fictitious satellite of Table\,\ref{tavola1}, $\omega$ stays essentially constant because of the frozen perigee configuration.  } in time and, thus, $\Phi(t)$ is quadratic. It is explicitly shown in  Fig.\,\ref{figura3} by the time series calculated for \rfr{Phi} from the same integration of the satellite's equations of motion.
\clearpage{}
It is an important feature because it allows to accurately separate the unwanted parabolic signature due to the atmospheric drag from the relativistic trend of interest affecting the time series of $\mathcal{M}(t)$ or $l(t)$, provided that a sufficiently long time span is chosen for the data analysis. The same holds, in principle, also for any other perturbing acceleration of non-gravitational origin inducing a secular trend in the satellite's semimajor axis like, e.g., the Yarkovsky-Rubincam thermal effect.
We numerically confirmed that by integrating the equations of motion of the fictitious satellite of Table\,\ref{tavola1} including the 1pN Schwarzschild-like and the atmospheric drag accelerations, and fitting a linear plus quadratic model to the resulting time series of $\Phi(t)$ over, say, 5 yr for a given value of $f_0$. As a result, we were able to accurately recover the slope of the relativistic secular signal. We successfully repeated it for different values of $f_0$ as well. It turns out that the longer the data span is, the more accurate the recovery of the linear signal. This suggests that, actually, also the mean anomaly $\mathcal{M}(t)$ and the mean longitude $l(t)$ may be fruitfully used in tests of pN gravity in the field of the Earth even with passive artificial satellites, contrary to the claims by Ciufolini and Pavlis \cite{2005NewA...10..636C}. The dependence of $\Phi(t)$ on $f_0$ may even represent an advantage to enhance the signal-to-noise ratio since, in principle, one can choose $f_0$ in order to maximize the relativistic rate for $\ang{\overset{\cdot}{\Phi}}$ to be added to the further contribution due to $\ang{\overset{\cdot}{\eta}},\,\ang{\overset{\cdot}{\epsilon}}$.
\subsubsection{The mean anomaly at epoch $\eta$ and the mean longitude at epoch $\epsilon$}
\begin{figure}[htb]
\centering
\centerline{
\vbox{
\begin{tabular}{c}
\epsfysize= 7.0 cm\epsfbox{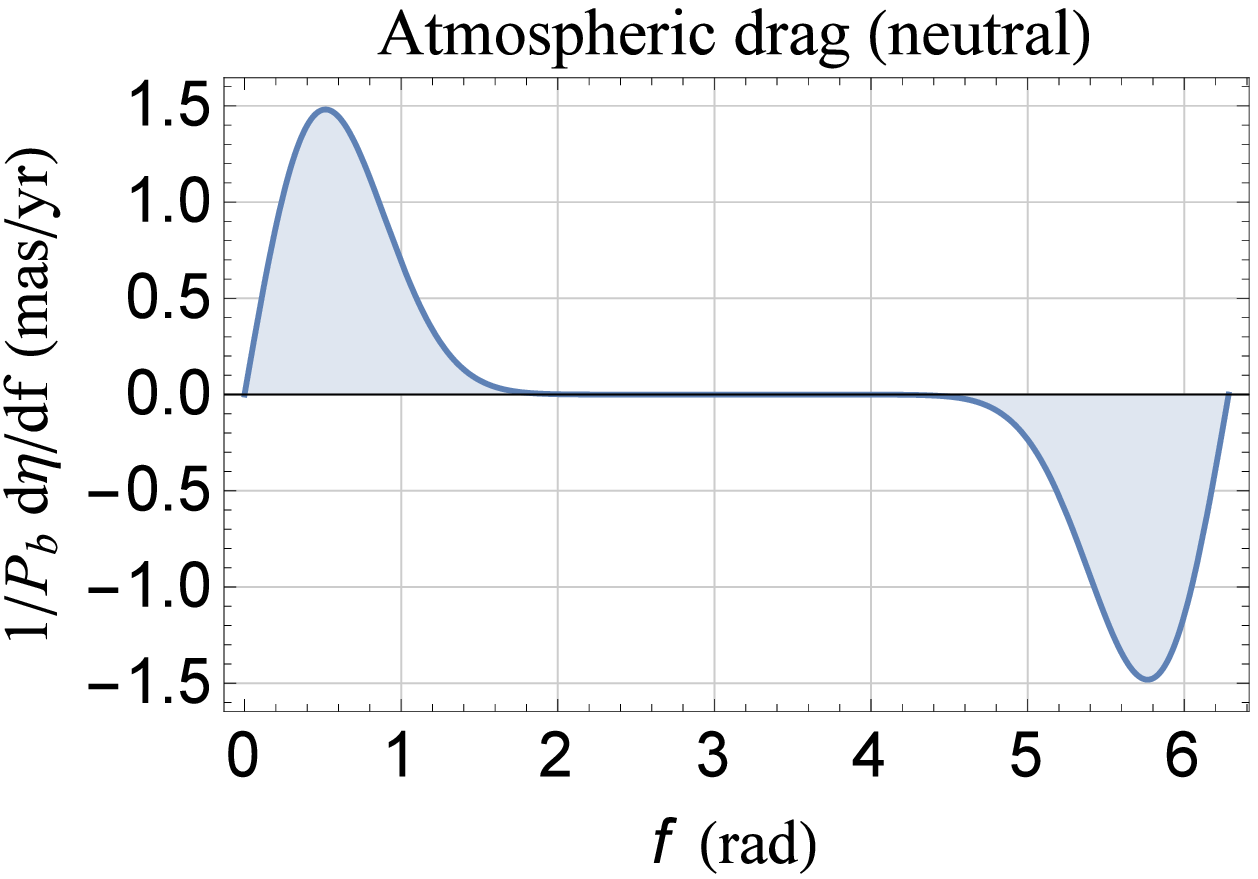}\\
\epsfysize= 7.0 cm\epsfbox{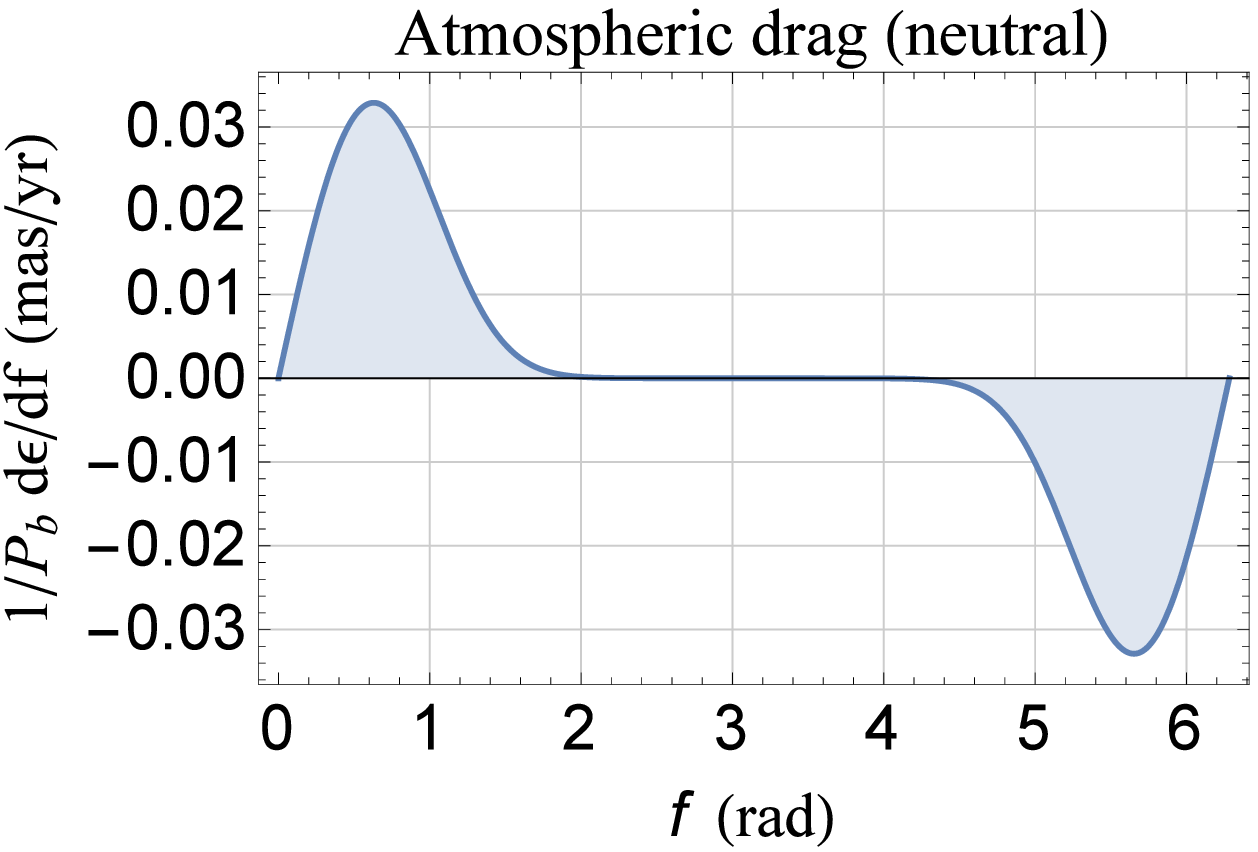}\\
\end{tabular}
}
}
\caption{Plots of \rfrs{pippo}{pluto}, computed for the orbital configuration of Table\,\ref{tavola1} and with $f_0 = 228\,\mathrm{deg}$, over a full orbital cycle of the true anomaly $f$. Their areas give $\ang{\overset{\cdot}{\eta}},\,\ang{\overset{\cdot}{\epsilon}}$ in mas yr$^{-1}$. In this case, they vanish, as confirmed also by a numerical integration of the satellite's equations of motion for the same physical and orbital parameters.}\label{figura4}
\end{figure}

\lb{drageteps}
About the secular rates of $\eta$ and $\epsilon$, the Gauss equations for their variations allow to obtain
\begin{align}
\rp{\nk}{2\uppi}\,\dert{\eta}{f} \nonumber \lb{pippo}&= \rp{C_\mathrm{D}\,\rho\ton{f}\,\Sigma\,\nk\,\mathcal{V}\ton{f}\,\ton{1-e^2}^2\sin f}{4\uppi\,e\,\ton{1 + e \cos f}^4}\times \\ \nonumber \\
\nonumber &\times \qua{2 + 3 e^2 + 2 e \ton{2 + e^2}\cos f + e^2 \cos 2f -\right. \\ \nonumber \\
&\left. -\nu\ton{1 - e^2}^{3/2} \ton{2 + e\cos f}\cos I}, \\ \nonumber \\
\rp{\nk}{2\uppi}\,\dert{\epsilon}{f} \nonumber \lb{pluto}& = -\rp{C_\mathrm{D}\,\rho\ton{f}\,\Sigma\,\nk\,\mathcal{V}\ton{f}\,\ton{1-e^2}}{8\uppi\,\ton{1+\sqrt{1-e^2}}\,\ton{1 + e \cos f}^4}\times \\ \nonumber \\
\nonumber &\times \grf{
4\,e\,(1 + e \cos f)\qua{-1 + e^2 \ton{1 + \sqrt{1-e^2}} + \right.\right.\\ \nonumber \\
\nonumber &\left.\left. + e\,\sqrt{1-e^2}\,\cos f}\,\sin f + \right.\\ \nonumber \\
\nonumber &\left. -\nu\ton{1-e^2}^2\,\qua{ \ton{1+\sqrt{1-e^2}}\ton{1-\cos I}\sin 2 u -\right.\right.\\ \nonumber \\
&\left.\left. -2\,e\,\cos I\,\ton{2 +e\,\cos f}\,\sin f }}.
\end{align}
Since it is not possible to analytically integrate \rfr{pippo} and \rfr{pluto} in an exact form, we, first, plot them as functions of $f$ over a full orbital cycle in Fig.\,\ref{figura4} for the orbital configuration of Table\,\ref{tavola1}, and, then, numerically calculate the areas under their curves in order to obtain $\ang{\overset{\cdot}{\eta}},\,\ang{\overset{\cdot}{\epsilon}}$.
\clearpage{}
Also in this case, a numerical integration of the satellite's equations of motion turns out to confirm such results.
\section{Some possible uses with the LAGEOS and LAGEOS II satellites}\lb{combos}
As an illustrative example, here we will look at the possibility of using the nodes $\Omega$ and the mean anomalies at epoch $\eta$ of, say, the existing satellites LAGEOS and LAGEOS II in order to propose an accurate test of the 1pN Lense-Thirring effect exploiting their multidecadal data records.

The availability of $\eta$ in addition to $\Omega$ may be particularly important in view of the fact that the competing classical secular precessions due to the even zonals of low degree, which have just the same time signature of the gravitomagnetic ones of interest, are nominally several orders of magnitude larger than them; thus, the signal-to-noise ratio must be somehow enhanced. The present-day level of actual mismodeling in the geopotential coefficients, which should be considered as (much) worse than the mere formal, statistical sigmas of the various global gravity field solutions\footnote{They are freely available on the Internet at the webpage of the International Centre for Global Earth Models (ICGEM), currently located at http://icgem.gfz-potsdam.de/tom$\_$longtime.} releasing the experimentally estimated values of the geopotential's parameters, does not yet allow to use the residuals of a single orbital element separately. To circumvent such an issue, some strategies involving the simultaneous use of more than one orbital element have been devised so far over the years: for a general overview, see, e.g., Renzetti \cite{2013CEJPh..11..531R}, and references therein.
To the benefit of the reader, we review here the linear combination approach, which is an extension of the one proposed by Ciufolini \cite{CiufoNCA96} to test the gravitomagnetic field of the Earth with artificial satellites of the LAGEOS family. \textcolor{black}{In turn, it is a generalization of the strategy put forth, for the first time, by I.I. Shapiro \cite{1990grg..conf..313S} who, at that time, wanted to separate the Sun-induced 1pN gravitoelectric perihelion precession from that due to the solar quadrupole mass moment $J_2$ by using other planets or highly eccentric asteroids.}

By looking at $N$ orbital elements\footnote{At least one of them must be affected also by the 1pN effect one is looking for. The $N$ orbital elements $\kappa^{(i)}$ may be different from one another belonging to the same satellite, or some of them may be identical belonging to different spacecraft (e.g., the nodes of two different vehicles).}
$\kappa^{(i)},\,i=1,2,\ldots N$  experiencing, among other things, classical secular precessions due to the even zonals of the geopotential, the following $N$ linear combinations can be written down
\eqi
\upmu_\mathrm{1pN}\,\ang{\overset{\cdot}{\kappa}}_\mathrm{1pN}^{(i)}+ \sum_{s=1}^{N-1}\,\derp{\ang{\overset{\cdot}{\kappa}}_{J_{2s}}^{(i)}}{J_{2s}}\,\delta J_{2s},\,i=1,2,\ldots N. \lb{lin}
\eqf
They involve the 1pN averaged precessions $\ang{\overset{\cdot}{\kappa}}_\mathrm{1pN}^{(i)}$ as predicted by General Relativity and scaled by a multiplicative parameter\footnote{It is equal to 1 in the Einstein's theory of gravitation, and 0 in the Newtonian one. In general, $\upmu_\mathrm{1pN}$ is not necessarily one of the parameters of the parameterized post-Newtonian (PPN) formalism, being possibly a combination of some of them.} $\upmu_\mathrm{1pN}$, and the errors in the computed  secular node precessions due to the uncertainties  in the first $N-1$ even zonals $J_{2s},~s=1,2,\ldots N-1$, assumed as mismodeled through $\delta J_{2s},~s=1,2,\ldots N-1$.
In the following and in Appendix\,\ref{precess}, we will use the shorthand
\eqi
\overset{\cdot}{\kappa}_{.\ell}\doteq\derp{\ang{\overset{\cdot}{\kappa}}_{J_\ell}}{J_\ell}
\eqf
for the partial derivative of the classical averaged precession $\ang{\overset{\cdot}{\kappa}}_{J_\ell}$ with respect to the generic even zonal $J_\ell$ of degree $\ell$. Then, the $N$ combinations of \rfr{lin} are posed equal to the experimental residuals $\delta\overset{\cdot}{\kappa}^{(i)},~i=1,2,\ldots N$ of each of the $N$ orbital elements considered getting
\eqi
\delta\overset{\cdot}{\kappa}^{(i)} = \upmu_\mathrm{1pN}\,\ang{\overset{\cdot}{\kappa}}_\mathrm{1pN}^{(i)}+ \sum_{s=1}^{N-1}\,\overset{\cdot}{\kappa}^{(i)}_{.2s}\,\delta J_{2s},~i=1,2,\ldots N. \lb{lineq} \eqf
It should be recalled that, in principle, the residuals $\delta\overset{\cdot}{\kappa}^{(i)}$ account for the purposely unmodelled 1pN effect, the mismodelling of the static and time-varying parts of the geopotential, and the non-gravitational forces.
If we look at the 1pN scaling parameter $\upmu_\mathrm{1pN}$ and the mismodeling in the even zonals $\delta J_{2s},~s=1,2,\ldots N-1$ as unknowns, we can interpret \rfr{lineq} as an inhomogenous linear system of $N$ algebraic equations in the $N$ unknowns
\eqi
\underbrace{\upmu_\mathrm{1pN},~\delta J_2,~\delta J_4 \ldots \delta J_{2(N-1)}}_{N},
\eqf
whose coefficients are
\eqi
\ang{\overset{\cdot}{\kappa}}^{(i)}_\mathrm{1pN},\,\overset{\cdot}{\kappa}^{(i)}_{.2s},\,i=1,2,\ldots N,\,s=1,2,\ldots N-1,
\eqf
while the constant terms are the $N$ orbital residuals
\eqi
\delta\overset{\cdot}{\kappa}^{(i)},\,i=1,2,\ldots N.
\eqf
It turns out that, after some algebraic manipulations,  the dimensionless 1pN scaling parameter can be expressed as
\eqi
\upmu_\mathrm{1pN}=\rp{\mathcal{C}_\delta}{\mathcal{C}_\mathrm{1pN}}.\lb{rs}
\eqf
In \rfr{rs}, the combination of the $N$ orbital residuals
\eqi
{\mathcal{C}}_\delta \doteq \delta\overset{\cdot}{\kappa}^{(1)} + \sum_{j=1}^{N-1}\,c_j\,\delta\overset{\cdot}{\kappa}^{(j+1)} \lb{combo}
\eqf
is, by construction, independent of the first $N-1$ even zonals, being, instead, impacted by the other ones of degree $\ell > 2(N-1)$ along with the non-gravitational perturbations and other possible orbital perturbations which cannot be reduced to the same formal expressions of the first $N-1$ even zonal rates. On the other hand,
\eqi
\mathcal{C}_\mathrm{1pN} \doteq \ang{\overset{\cdot}{\kappa}}_\mathrm{1pN}^{(1)} + \sum_{j=1}^{N-1}\,c_j\,\ang{\overset{\cdot}{\kappa}}_\mathrm{1pN}^{(j+1)} \lb{combopN}\eqf
combines the $N$ 1pN orbital precessions as predicted by General Relativity.
The dimensionless coefficients $c_j,\ j=1,2,\ldots N-1$ in \rfr{combo}-\rfr{combopN} depend only on some of the orbital parameters of the satellite(s) involved in such a way that, by construction, ${\mathcal{C}}_\delta=0$ if \rfr{combo} is calculated by posing
\eqi
\delta\overset{\cdot}{\kappa}^{(i)}=\overset{\cdot}{\kappa}^{(i)}_{.\ell}\,\delta J_{\ell},\ i=1,2,\ldots N
\eqf
for any of the first $N-1$ even zonals, independently of the value assumed for its uncertainty $\delta J_{\ell}$.

As far as the Lense-Thirring effect and the satellites LAGEOS and LAGEOS II are concerned, the linear combination of the four experimental residuals $\delta\Omega^\mathrm{L},\,\delta\Omega^\mathrm{L\,II},\,\delta\eta^\mathrm{L},\,\delta\eta^\mathrm{L\,II}$ of the satellites's nodes and mean anomalies at epoch suitably designed to cancel out the secular precessions due to the first three even zonal harmonics $J_2,\,J_4,\,J_6$ of the geopotential is
\eqi
\mathcal{C}_\delta=\delta\Omega^\mathrm{L}+c_1\,\delta\Omega^\mathrm{L\,II}+c_2\,\delta\eta^\mathrm{L}+c_3\,\delta\eta^\mathrm{L\,II}\lb{kombo}
\eqf
whose coefficients $c_1,~c_2,~c_3$ are purposely constructed with the results of Section~\ref{secu}. They turn out to be
\begin{align}
D\,c_1 \lb{c1}\nonumber & = \nddL\,\etqL\,\etsLII -\etdL\,\ndqL\,\etsLII -\nddL\,\etqLII\,\etsL +\\ \nonumber \\
&+ \etdLII\,\ndqL\,\etsL +\etdL\,\etqLII\,\ndsL -\etdLII\,\etqL\,\ndsL,\\\nonumber\\
D\,c_2 \lb{c2} \nonumber &= -\nddL\,\ndqLII\,\etsLII +\nddLII\,\ndqL\,\etsLII +\nddL\,\etqLII\,\ndsLII -\\ \nonumber \\
&-\etdLII\,\ndqL\,\ndsLII -\nddLII\,\etqLII\,\ndsL +\etdLII\,\ndqLII\,\ndsL,\\\nonumber\\
D\,c_3 \lb{c3} \nonumber &= -\nddL\,\etqL\,\ndsLII +\etdL\,\ndqL\,\ndsLII +\nddL\,\ndqLII\,\etsL -\\ \nonumber \\
&- \nddLII\,\ndqL\,\etsL -\etdL\,\ndqLII\,\ndsL +\nddLII\,\etqL\,\ndsL,
\end{align}
where the common denominator is
\begin{align}
D\nonumber &= \etdL\,\ndqLII\,\etsLII -\nddLII\,\etqL\,\etsLII -\etdL\,\etqLII\,\ndsLII +\\ \nonumber \\
& + \etdLII\,\etqL\,\ndsLII +\nddLII\,\etqLII\,\etsL -\etdLII\,\ndqLII\,\etsL.
\end{align}
Their numerical values, computed with the satellites' orbital elements inserted in \rfrs{OJ2}{etaJ6}, are
\begin{align}
c_1 \lb{C1}& = 2.77536, \\ \nonumber \\
c_2 \lb{C2}& = -2.46439, \\ \nonumber \\
c_3 \lb{C3}& = 10.9532.
\end{align}
Thus, the predicted combined Lense-Thirring signature is
\begin{align}
\mathcal{C}_\mathrm{LT} \nonumber &= \overset{\cdot}{\Omega}_\textrm{LT}^\textrm{L} + c_1\,\overset{\cdot}{\Omega}_\textrm{LT}^\textrm{L\,II} + c_2\,\overset{\cdot}{\eta}_\textrm{LT}^\textrm{L} + c_3\,\overset{\cdot}{\eta}_\textrm{LT}^\textrm{L\,II} = \\ \nonumber \\
& = 118.04~\textrm{mas~yr}^{-1}.\lb{comboLT}
\end{align}
The combination of \rfr{kombo} is mainly affected by the orbital precessions induced by  the fourth even zonal harmonic $J_8$ of the geopotential. The resulting mismodeled combined signal can be evaluated by means of \rfrs{OJ8}{etaJ8} along with some measure of the uncertainty in $J_8$.
If one were to rely upon on the formal sigmas of the latest global Earth's gravity field models by the dedicated GRACE and GOCE missions, the resulting impact on \rfr{comboLT} would be much smaller than $1\%$. Indeed, from, e.g., the zero-tide model Tongji-Grace02s \cite{2018JGRB..123.6111C}, it is\footnote{The zonal harmonics $J_\ell$ of the geopotential are connected with its fully normalized Stokes coefficients ${\overline{C}}_{\ell,0}$ by the relation $J_\ell = -\sqrt{2\ell+1}\,{\overline{C}}_{\ell,0},\,\ell=2,\,3,\,4,\ldots$}
$\upsigma_{{\overline{{C}}_{8,0}}} = 1.3\times 10^{-14}$.
It implies a combined mismodeled precessions as little as $0.01\,\mathrm{mas\,yr}^{-1}$,
corresponding to $0.01\%$ of the combined Lense-Thirring effect.
If, instead, the difference $\Delta {\overline{C}}_{8,0}$ between the values of ${\overline{C}}_{8,0}$ from Tongji-Grace02s and the zero-tide model ITU$\_$GRACE16 \cite{ITU16}, whose formal errors are comparable, is adopted as a measure of the actual uncertainty in the even zonal of degree 8, the resulting mismodeled signal amounts to $2.1\,\mathrm{mas\,yr}^{-1}$  corresponding to a percent error in the Lense-Thirring combined signature of $1.8\%$.

In fact, an accurate investigation, both analytical and numerical, of the perturbations on $\eta$ induced by the main non-gravitational accelerations acting on the LAGEOS-type satellites like, e.g., the direct solar radiation pressure, the Earth's albedo, the Earth's direct infrared radiation pressure, the Earth's Yarkovsky-Rubincam and Solar Yarkovsky-Schach thermal effects, possible anisotropic reflectivity, etc. \cite{2001P&SS...49..447L,2002P&SS...50.1067L,2003GeoRL..30.1957L,2017AcAau.140..469P,2018PhRvD..98d4034V,Lucchesi019} is required to realistically assess the overall error budget of the promising combination of \rfr{kombo}. This is outside the scopes of the present paper.
\section{Summary and overview}\lb{fine}
In presence of Newtonian, general relativistic 1pN or modified gravity-induced disturbing accelerations, the shifts $\Delta\mathcal{M}(t)$ and $\Delta l(t)$ of the mean anomaly $\mathcal{M}(t)$ and the mean longitude $l(t)$ with respect to their Keplerian linear trends are, in general, due to the perturbations $\Delta\eta(t)$ and $\Delta\epsilon(t)$ of the mean anomaly at epoch $\eta$ and mean longitude at epoch $\epsilon$, and the change $\Delta\nk(t)$ in the mean motion $\nk$ which, in some cases, can induce a quadratic shift $\Phi(t)$ in $\mathcal{M}(t)$ and $l(t)$ depending on the true anomaly at epoch $f_0$.

In the case of an Earth's artificial satellite, the atmospheric drag affects $\Phi(t)$ quadratically; nonetheless, the non-Newtonian linear trends of interest may be effectively separated from such a potentially competing aliasing effect if a sufficiently long time span for the data analysis is adopted. Thus, also $\mathcal{M}(t)$ and $l(t)$ can, in principle, be employed in  gravity tests even with passive geodetic satellites, not to mention the use of drag-free apparatuses. If, instead, $\eta$ and $\epsilon$ are adopted, such an issue is a-priori circumvented because they are not impacted by the possible change in the mean motion $\nk$. Since $\eta$ and $\epsilon$ undergo secular precessions due to the even zonal harmonics $J_\ell,\,\ell=2,\,4,\ldots$ of the geopotential, it is possible, in principle, to use them in combination with, say, the nodes $\Omega$ to reduce the impact of the mismodeled even zonals in experiments of fundamental physics with existing satellites. In an actual test, a detailed analysis of the perturbations affecting $\eta$ and $\epsilon$ by all the most relevant non-gravitational accelerations  should be performed. There are no net Lense-Thirring rates of change of the semimajor axis  $a$ and $\nk$.

In astronomical binary systems, not affected by non-gravitational perturbations, using $\eta$ may provide a further valuable observable in addition to the usual periastron precession to put to the test general relativity and, say, modified models of gravity, or to better characterize the physical properties of the bodies like, e.g., their oblateness $J_2$ and their orbital configurations as well. Indeed, the 1pN effects on $\eta$ are often larger than the corresponding pericenter rates.
\begin{appendices}
\section{Mean orbital precessions of $\Omega$ and $\eta$ due to the even zonal harmonics of the geopotential}\lb{precess}
\renewcommand{\theequation}{A\arabic{equation}}
\setcounter{equation}{0}
Here, we analytically calculate the coefficients
\eqi
\overset{\cdot}{\kappa}_{.\ell}\doteq\derp{\ang{\overset{\cdot}{\kappa}}_{J_\ell}}{J_{\ell}},~\ell = 2,\,4,\,6,\,8,\,\kappa = \Omega,\,\eta
\eqf
of the precessions
\eqi
\ang{\overset{\cdot}{\kappa}}_{J_\ell},~\ell = 2,\,4,\,6,\,8,\,\kappa = \Omega,\,\eta,
\eqf
of the node $\Omega$ and of the mean anomaly at epoch $\eta$
averaged over one full orbital period $\Pb$, induced by the first four even zonal harmonics $J_{\ell}$.
To this aim, we use the standard Lagrange planetary equations \cite{Bertotti03}
\begin{align}
\ang{\overset{\cdot}{\Omega}}  &= - \rp{1}{\nk\,a^2\,\sin I\,\sqrt{1 - e^2}}\derp{\ang{\Delta U_{\ell}}}{I}, \\ \nonumber \\
\ang{\overset{\cdot}{\eta}} & = \rp{2}{\nk\,a}\derp{\ang{\Delta U_\ell}}{a} +\rp{\ton{1-e^2}}{\nk\,a^2\,e}\derp{\ang{\Delta U_\ell}}{e}.
\end{align}
In them, the correction of degree $\ell$
\eqi
\Delta U_\ell\ton{\bds r} = \rp{\mu}{r}\,\ton{\rp{R_\mathrm{e}}{r}}^\ell\,J_\ell\,\mathcal{P}_\ell\ton{\xi},\,\ell=2,\,4,\ldots 8\lb{kuzz}
\eqf
to the Newtonian monopole is straightforwardly averaged over one full orbital revolution by using the Keplerian ellipse
%\begin{align}
%r = \rp{a\ton{1-e^2}}{1 + e\cos f}, \\ \nonumber \\
%
%\mathrm{d}t = \rp{\ton{1 - e^2}^{3/2}}{\nk\,\ton{1 + e \cos f}^2}\,\mathrm{d}f.
%\end{align}
as reference unperturbed orbit. In \rfr{kuzz}, $\mathcal{P}_\ell\ton{\xi}$ is the Legendre polynomial of degree $\ell$.
As a result, two kind of averaged, long-term effects occur: secular precessions, explicitly displayed in Section~\ref{secu} and labelled with a superscript \virg{s}, and long-periodic signatures, not shown here, having a harmonic pattern characterized by a frequency which is an integer multiple of that of  perigee $\omega$. In the calculation, the Earth's symmetry axis $\bds{\hat{S}}$ is assumed to be aligned with the reference $z$ axis; moreover, no a-priori simplifying assumptions concerning the orbital geometry of the satellite were made at all.
\subsection{Secular effects}\lb{secu}
\begin{align}
\overset{\cdot}{\Omega}^\textrm{s}_{.2} \lb{OJ2}& = -\rp{3\,\nk\,R_\mathrm{e}^2\,\cos I}{2\,a^2\,\ton{1-e^2}^2}, \\ \nonumber \\
\overset{\cdot}{\eta}^\textrm{s}_{.2} \lb{EtaJ2}& = \rp{3\,\nk\,R_\mathrm{e}^2\,\ton{1 + 3\,\cos 2  I }}{8\,a^2\,\ton{1 - e^2}^{3/2}}, \\ \nonumber \\
\overset{\cdot}{\Omega}^\textrm{s}_{.4} \lb{OJ4}& = \rp{15\,\nk\,R_\mathrm{e}^4\,\ton{2 + 3\,e^2}\ton{9\,\cos I + 7\,\cos 3I}}{128\,a^4\,\ton{1-e^2}^4}, \\ \nonumber \\
\overset{\cdot}{\eta}^\textrm{s}_{.4} \lb{etaJ4}& = -\rp{45\,\nk\,R_\mathrm{e}^4\,e^2\,\ton{9 + 20\,\cos 2  I  + 35\,\cos 4  I }}{1,024\,a^4\,\ton{1 - e^2}^{7/2}}, \\ \nonumber \\
\overset{\cdot}{\Omega}^\textrm{s}_{.6} \lb{OJ6}& = -\rp{105\,\nk\,R_\mathrm{e}^6\,\ton{8 + 40\,e^2 + 15\,e^4}\,\ton{50\,\cos I + 45\,\cos 3I + 33\,\cos 5I}}{16,384\,a^6\,\ton{1-e^2}^6}, \\ \nonumber \\
\overset{\cdot}{\eta}^\textrm{s}_{.6} \nonumber \lb{etaJ6} & = \rp{35\,\nk\,R_\mathrm{e}^6}{65,536\,a^6\,\ton{1 - e^2}^{11/2}}\,\ton{-8 + 20\,e^2 +
15\,e^4}\times\\ \nonumber \\
&\times\ton{50 + 105\,\cos 2  I  + 126\,\cos 4  I  + 231\,\cos 6  I }, \\ \nonumber\\
\overset{\cdot}{\Omega}^\textrm{s}_{.8} \nonumber \lb{OJ8}& =\rp{315\,\nk\,R_\mathrm{e}^8}{2,097,152\,a^8\,\ton{1 - e^2}^8}\,\grf{16 +7\,e^2 \qua{24 + 5\,e^2 \ton{6 + e^2}}}\times\\ \nonumber \\
&\times \qua{1,225\,\cos  I  + 11\,\ton{105\,\cos 3  I  + 91\,\cos 5  I  + 65\,\cos 7  I }}, \\ \nonumber\\
\overset{\cdot}{\eta}^\textrm{s}_{.8} \nonumber \lb{etaJ8}& =    -\rp{315\,\nk\,R_\mathrm{e}^8}{33,554,432\,a^8\,\ton{1 - e^2}^{15/2}}\,\qua{-32 + 35\,e^4 \ton{4 + e^2}}\times\\ \nonumber \\
&\times \ton{1,225 +2,520\,\cos 2  I  + 2,772\,\cos 4 I  + 3,432\,\cos 6  I  + 6,435\,\cos 8  I }.
\end{align}
\end{appendices}
\bibliography{MS_binary_pulsar_bib,Gclockbib,semimabib,PXbib}{}

\end{document}